\documentclass[a4paper,12pt]{article}
\usepackage{Style}

\onehalfspace

\title
{ %
\vspace*{5.0cm} \LARGE{\bf Single-Phase Flow of Non-Newtonian Fluids in Porous Media} \vspace*{4.0cm} \\
}

\author{Taha Sochi\footnote{University College London - Department of Physics \& Astronomy - Gower Street - London - WC1E 6BT. Email:
t.sochi@ucl.ac.uk.} \vspace*{3.0cm}}

\date{2009}

\makeindex

\begin{document}

\maketitle %
\pagenumbering{roman}

\newpage
\phantomsection \addcontentsline{toc}{section}{Contents} %
\tableofcontents

\newpage
\phantomsection \addcontentsline{toc}{section}{List of Figures} %
\listoffigures

\phantomsection \addcontentsline{toc}{section}{List of Tables} %
\listoftables


{\setlength{\parskip}{6pt plus 2pt minus 1pt}

\pagestyle{headings} %
\addtolength{\headheight}{+1.6pt}
\lhead[{Chapter \thechapter \thepage}]%
      {{\bfseries\rightmark}}
\rhead[{\bfseries\leftmark}]%
     {{\bfseries\thepage}} %
\headsep = 1.0cm               %

\newpage

\pagenumbering{arabic}

\phantomsection \addcontentsline{toc}{section}{Abstract} %
\noindent
{\large \bf Abstract} \vspace{0.5cm} \\
The study of flow of \nNEW\ fluids in porous media is very important and serves a wide variety of
practical applications in processes such as enhanced oil recovery from underground reservoirs,
filtration of polymer solutions and soil remediation through the removal of liquid pollutants.
These fluids occur in diverse natural and synthetic forms and can be regarded as the rule rather
than the exception. They show very complex strain and time dependent behavior and may have initial
\yields. Their common feature is that they do not obey the simple \NEW\ relation of proportionality
between stress and rate of deformation. \NNEW\ fluids are generally classified into three main
categories: \timeind\ whose strain rate solely depends on the instantaneous stress, \timedep\ whose
strain rate is a function of both magnitude and duration of the applied stress and viscoelastic
which shows partial elastic recovery on removal of the deforming stress and usually demonstrates
both time and strain dependency. In this article the key aspects of these fluids are reviewed with
particular emphasis on single-phase flow through porous media. The four main approaches for
describing the flow in porous media are examined and assessed. These are: continuum models, bundle
of tubes models, numerical methods and pore-scale network modeling.

\newpage

\section{Introduction} \label{Introduction}
\NEW\ fluids are defined to be those fluids exhibiting a direct proportionality between stress
$\sS$ and strain rate $\sR$ in laminar flow, that is
\begin{equation}\label{}
    \sS = \Vis \sR
\end{equation}
where the viscosity $\Vis$ is independent of the strain rate although it might be affected by other
physical parameters, such as temperature and pressure, for a given fluid system. A stress versus
strain rate graph for a \NEW\ fluid will be a straight line through the origin
\cite{Skellandbook1967, ChhabrabookR1999}. In more precise technical terms, \NEW\ fluids are
characterized by the assumption that the extra stress tensor, which is the part of the total stress
tensor that represents the shear and extensional stresses caused by the flow excluding hydrostatic
pressure, is a linear isotropic function of the components of the velocity gradient, and therefore
exhibits a linear relationship between stress and the rate of strain \cite{OwensbookP2002,
OsP2004}. In tensor form, which takes into account both shear and extension flow components, this
linear relationship is expressed by
\begin{equation}\label{NewtonianTensForm}
    \sTen = \Vis \rsTen
\end{equation}
where $\sTen$ is the extra stress tensor and $\rsTen$ is the rate of strain tensor which describes
the rate at which neighboring particles move with respect to each other independent of superposed
rigid rotations. \NEW\ fluids are generally featured by having shear- and \timeind\ viscosity, zero
normal stress differences in simple shear flow, and simple proportionality between the viscosities
in different types of deformation \cite{BirdbookAH1987, BarnesbookHW1993}.

\vspace{0.2cm}

All those fluids for which the proportionality between stress and strain rate is violated, due to
nonlinearity or initial \yields, are said to be \nNEW. Some of the most characteristic features of
\nNEW\ behavior are: strain-dependent viscosity as the viscosity depends on the type and rate of
deformation, \timedep\ viscosity as the viscosity depends on duration of deformation, \yields\
where a certain amount of stress should be reached before the flow starts, and stress relaxation
where the resistance force on stretching a fluid element will first rise sharply then decay with a
characteristic relaxation time.

\vspace{0.2cm}

\NNEW\ fluids are commonly divided into three broad groups, although in reality these
classifications are often by no means distinct or sharply defined \cite{Skellandbook1967,
ChhabrabookR1999}:
\begin{enumerate}

\item \Timeind\ fluids are those for which the strain rate at a given
point is solely dependent upon the instantaneous stress at that point.

\item \Vc\ fluids are those that show partial elastic recovery upon the
removal of a deforming stress. Such materials possess properties of both viscous fluids and elastic
solids.

\item \Timedep\ fluids are those for which the strain rate is a function
of both the magnitude and the duration of stress and possibly of the time lapse between consecutive
applications of stress.

\end{enumerate}

Those fluids that exhibit a combination of properties from more than one of the above groups are
described as complex fluids \cite{Collyer1974}, though this term may be used for \nNEW\ fluids in
general.

\vspace{0.2cm}

The generic rheological behavior of the three groups of \nNEW\ fluids is graphically presented in
Figures (\ref{TimeIndependent}-\ref{TimeDependent}). Figure (\ref{TimeIndependent}) demonstrates
the six principal rheological classes of the \timeind\ fluids in shear flow. These represent
\shThin, \shThik\ and shear-independent fluids each with and without \yields. It is worth noting
that these rheological classes are idealizations as the rheology of these fluids is generally more
complex and they can behave differently under various deformation and ambient conditions. However,
these basic rheological trends can describe the behavior under particular circumstances and the
overall behavior consists of a combination of stages each modeled with one of these basic classes.

\vspace{0.2cm}

Figures (\ref{VERheology1}-\ref{VERheology3}) display several aspects of the rheology of \vc\
fluids in bulk and \insitu. In Figure (\ref{VERheology1}) a stress versus time graph reveals a
distinctive feature of time-dependency largely observed in \vc\ fluids. As seen, the overshoot
observed on applying a sudden deformation cycle relaxes eventually to the equilibrium \steadys.
This \timedep\ behavior has an impact not only on the flow development in time, but also on the
\dilatancy\ behavior observed in porous media flow under \steadys\ conditions. The \convdiv\
geometry contributes to the observed increase in viscosity when the relaxation time characterizing
the fluid becomes comparable in size to the characteristic time of the flow, as will be discussed
in Section (\ref{Viscoelasticity}).

\vspace{0.2cm}

Figure (\ref{VERheology2}) reveals another characteristic feature of \vy\ observed in porous media
flow. The intermediate plateau may be attributed to the \timedep\ nature of the \vc\ fluid when the
relaxation time of the fluid and the characteristic time of the flow become comparable in size.
This behavior may also be attributed to build-up and break-down due to sudden change in radius and
hence rate of strain on passing through the \convdiv\ pores. A \thixotropic\ nature will be more
appropriate in this case.

\vspace{0.2cm}

Figure (\ref{VERheology3}) presents another typical feature of \vc\ fluids. In addition to the
low-deformation \NEW\ plateau and the \shThin\ region which are widely observed in many \timeind\
fluids and modeled by various \timeind\ rheological models, there is a thickening region which is
believed to be originating from the dominance of extension over shear at high flow rates. This
feature is mainly observed in the flow through porous media, and the \convdiv\ geometry is usually
given as an explanation for the shift in dominance from shear to extension at high flow rates.

\vspace{0.2cm}

In Figure (\ref{TimeDependent}) the two basic classes of \timedep\ fluids are presented and
compared to the \timeind\ fluid in a graph of stress versus time of deformation under constant
strain rate condition. As seen, \thixotropy\ is the equivalent in time to \shThin, while \rheopexy\
is the equivalent in time to \shThik.

\vspace{0.2cm}

A large number of models have been proposed in the literature to model all types of \nNEW\ fluids
under diverse flow conditions. However, the majority of these models are basically empirical in
nature and arise from curve-fitting exercises \cite{BarnesbookHW1993}. In the following sections,
the three groups of \nNEW\ fluids will be investigated and a few prominent examples of each group
will be presented.


\begin{figure}[!h]
  \centering{}
  \includegraphics
  [scale=0.6]
  {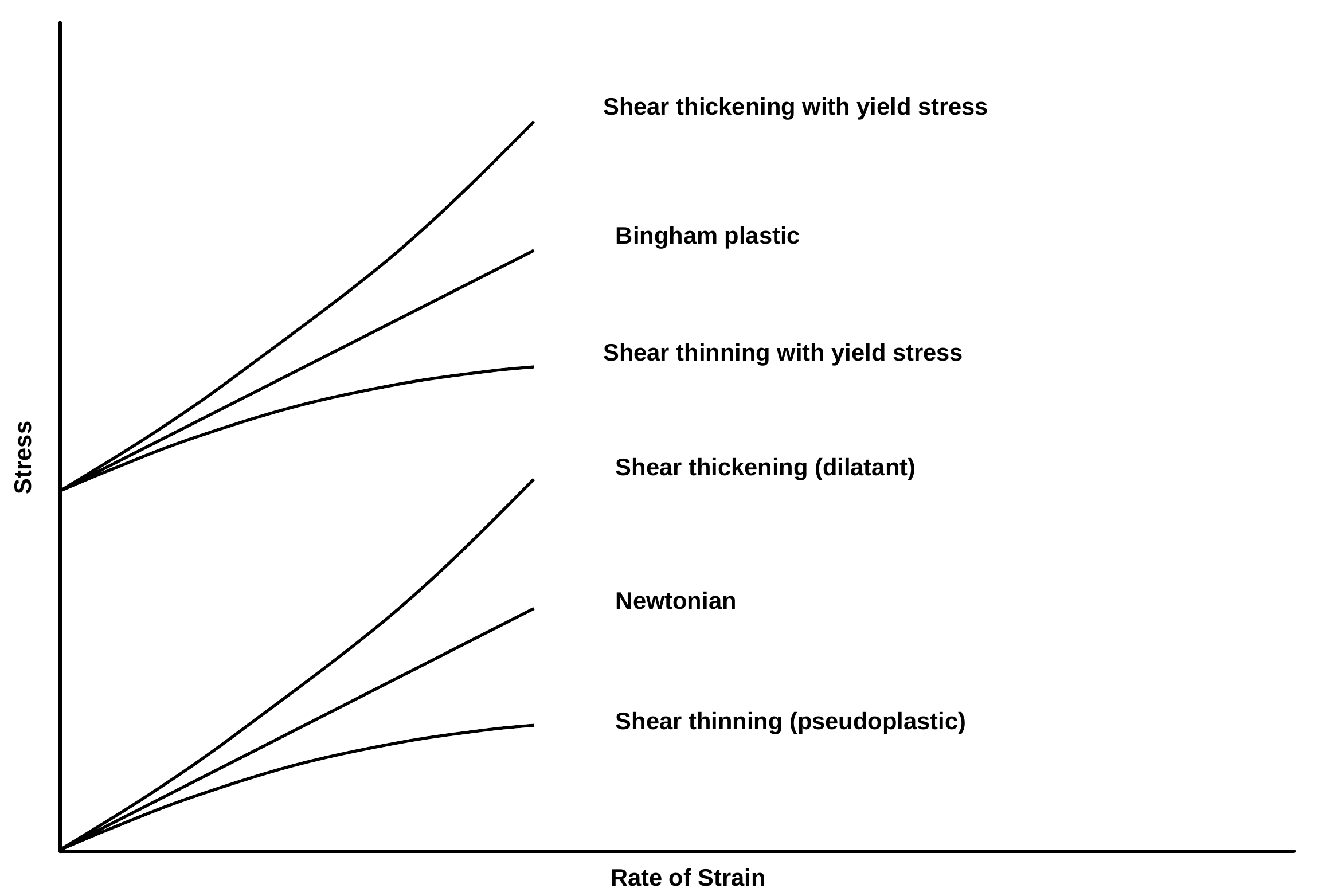}
  \caption[The six main classes of the \timeind\ fluids presented in
  a generic graph of stress against strain rate in shear flow]
  {The six main classes of the \timeind\ fluids presented in
  a generic graph of stress against strain rate in shear flow.}
  \label{TimeIndependent}
\end{figure}


\begin{figure}[!h]
  \centering{}
  \includegraphics
  [scale=0.57]
  {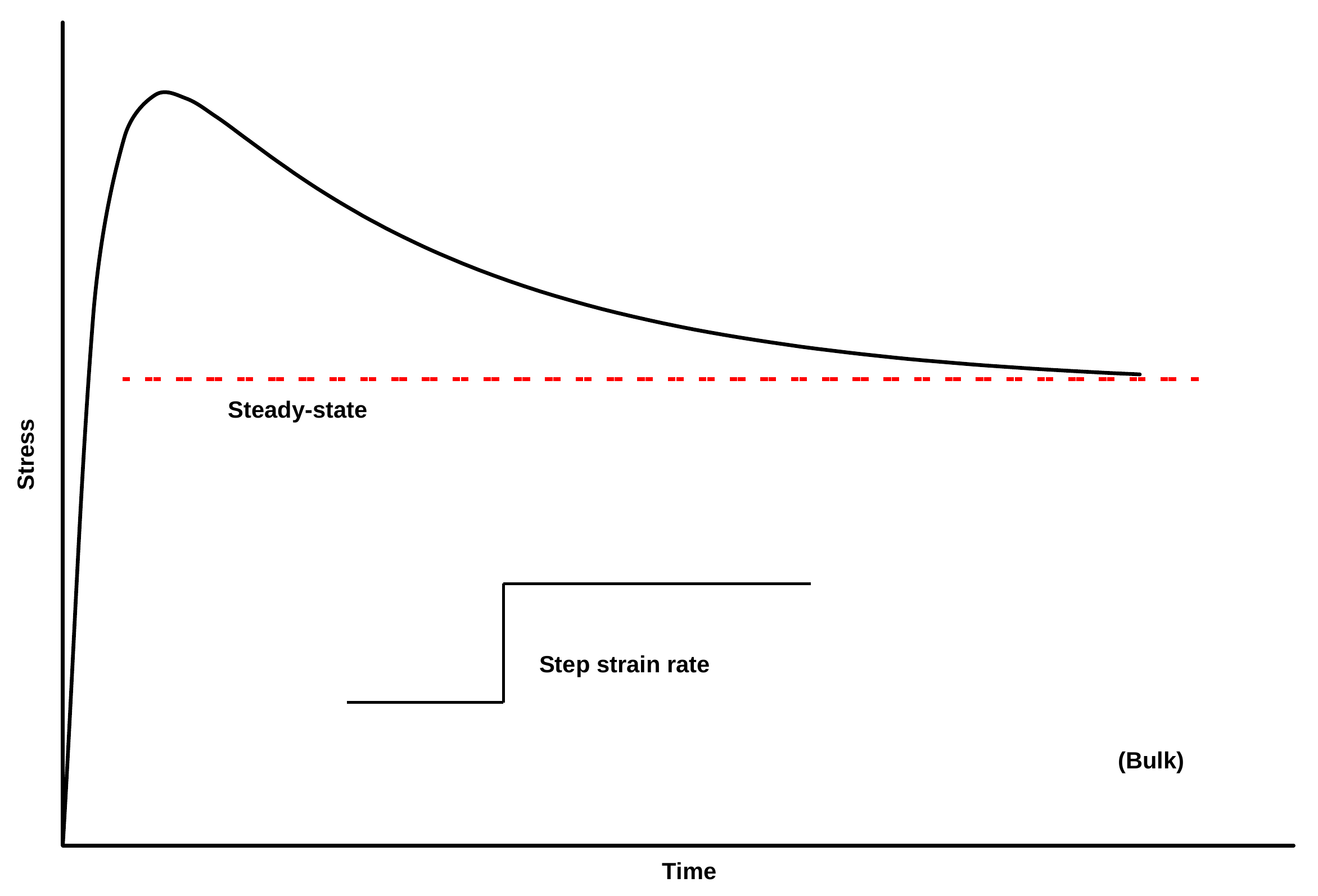}
  \caption[Typical time-dependence behavior of \vc\ fluids due to delayed response and relaxation
           following a step increase in strain rate]
  {Typical time-dependence behavior of \vc\ fluids due to delayed response and relaxation
   following a step increase in strain rate.}
  \label{VERheology1}
\end{figure}


\begin{figure}[!h]
  \centering{}
  \includegraphics
  [scale=0.57]
  {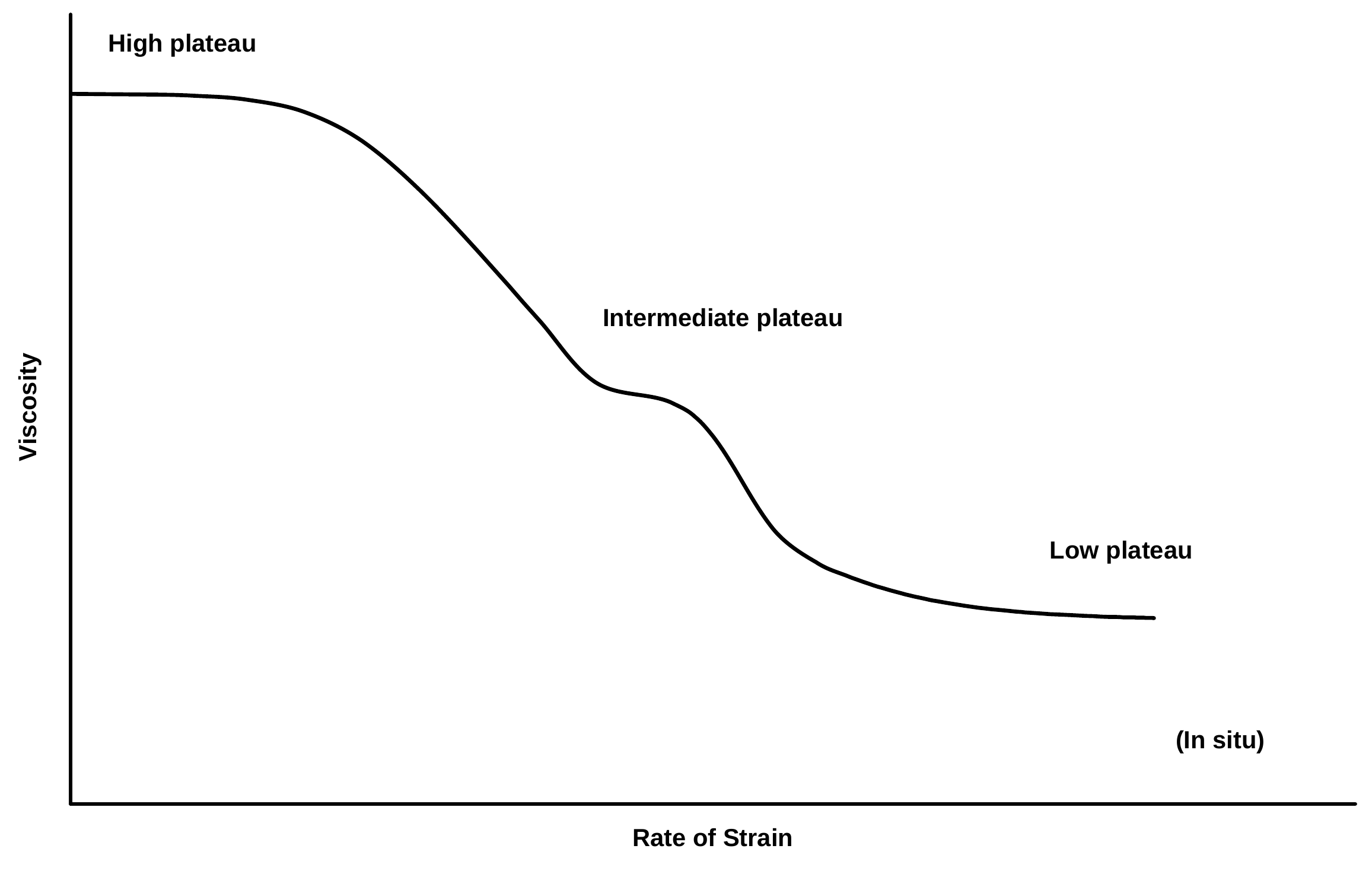}
  \caption[Intermediate plateau typical of \insitu\ \vc\ behavior due to time characteristics
  of the fluid and flow system and \convdiv\ nature of the flow channels]
  {Intermediate plateau typical of \insitu\ \vc\ behavior due to time characteristics
  of the fluid and flow system and \convdiv\ nature of the flow channels.}
  \label{VERheology2}
\end{figure}


\begin{figure}[!h]
  \centering{}
  \includegraphics
  [scale=0.57]
  {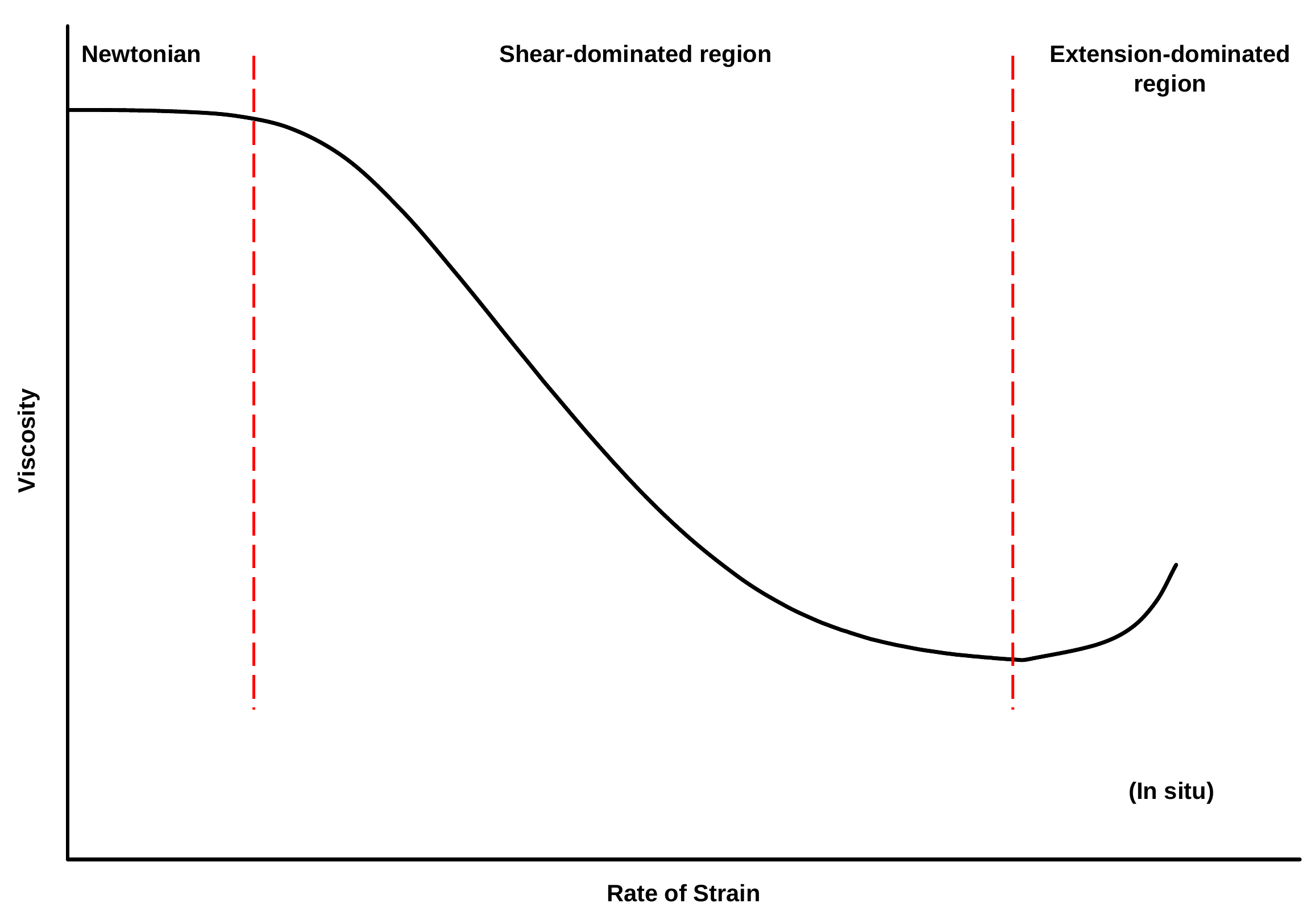}
  \caption[Strain hardening at high strain rates typical of \insitu\ \vc\ flow
  attributed to the dominance of extension over shear at high flow rates]
  {Strain hardening at high strain rates typical of \insitu\ \vc\ flow
  attributed to the dominance of extension over shear at high flow rates.}
  \label{VERheology3}
\end{figure}


\begin{figure}[!h]
  \centering{}
  \includegraphics
  [scale=0.57]
  {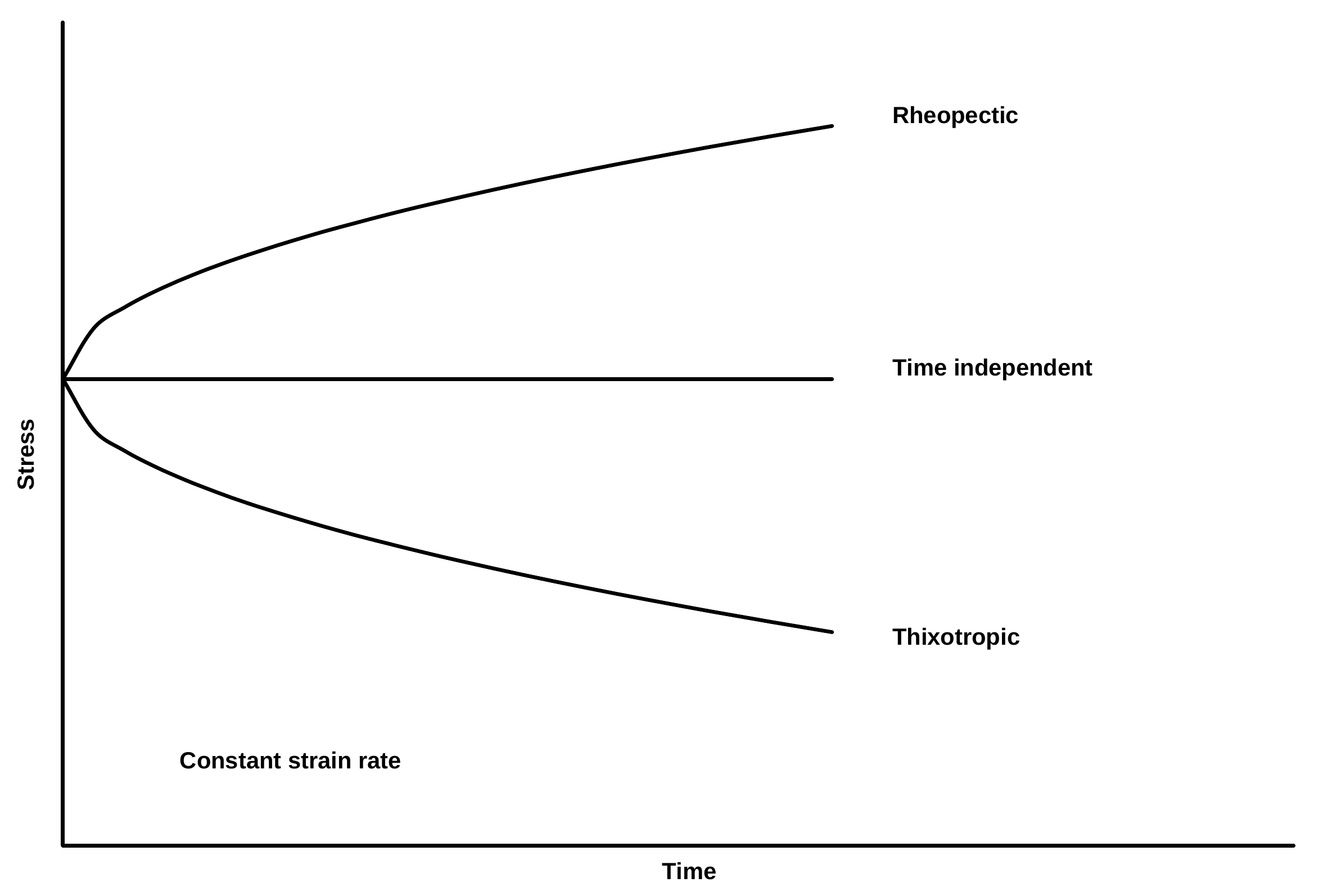}
  \caption[The two classes of \timedep\ fluids compared to the \timeind\
           presented in a generic graph of stress against time]
  {The two classes of \timedep\ fluids compared to the \timeind\
  presented in a generic graph of stress against time.}
  \label{TimeDependent}
\end{figure}

\newpage

\subsection{\TimeInd\ Fluids} \label{}
Shear rate dependence is one of the most important and defining characteristics of \nNEW\ fluids in
general and \timeind\ fluids in particular. When a typical \nNEW\ fluid experiences a shear flow
the viscosity appears to be \NEW\ at low shear rates. After this initial \NEW\ plateau the
viscosity is found to vary with increasing shear rate. The fluid is described as \shThin\ or
\pseudoplastic\ if the viscosity decreases, and \shThik\ or \dilatant\ if the viscosity increases
on increasing shear rate. After this shear-dependent regime, the viscosity reaches a limiting
constant value at high shear rate. This region is described as the upper \NEW\ plateau. If the
fluid sustains initial stress without flowing, it is called a \yields\ fluid. Almost all polymer
solutions that exhibit a shear rate dependent viscosity are \shThin, with relatively few polymer
solutions demonstrating \dilatant\ behavior. Moreover, in most known cases of \shThik\ there is a
region of \shThin\ at lower shear rates \cite{BirdbookAH1987, BarnesbookHW1993, OwensbookP2002}.

\vspace{0.2cm}

In this article, we present four fluid models of the \timeind\ group: \powlaw, \Ellis, \Carreau\
and \HB. These are widely used in modeling \nNEW\ fluids of this group.

\subsubsection{\PowLaw\ Model} \label{powLawSec}
The \powlaw, or \OstWae\ model, is one of the simplest \timeind\ fluid models as it contains only
two parameters. The model is given by the relation \cite{BirdbookAH1987, CarreaubookKC1997}
\begin{equation}\label{powLawEq}
    \Vis = C \sR^{n-1}
\end{equation}
where $\Vis$ is the viscosity, $C$ is the consistency factor, $\sR$ is the shear rate and $n$ is
the flow behavior index. In Figure (\ref{BulkRheoPowLaw}), the bulk rheology of this model for
\shThin\ case is presented in a generic form as viscosity versus shear rate on log-log scales. The
\powlaw\ is usually used to model \shThin\ though it can also be used for modeling \shThik\ by
making $n>1$. The major weakness of \powlaw\ model is the absence of plateaux at low and high shear
rates. Consequently, it fails to produce sensible results in these shear regimes.


\begin{figure} [!h]
  \centering{}
  \includegraphics
  [scale=0.6]
  {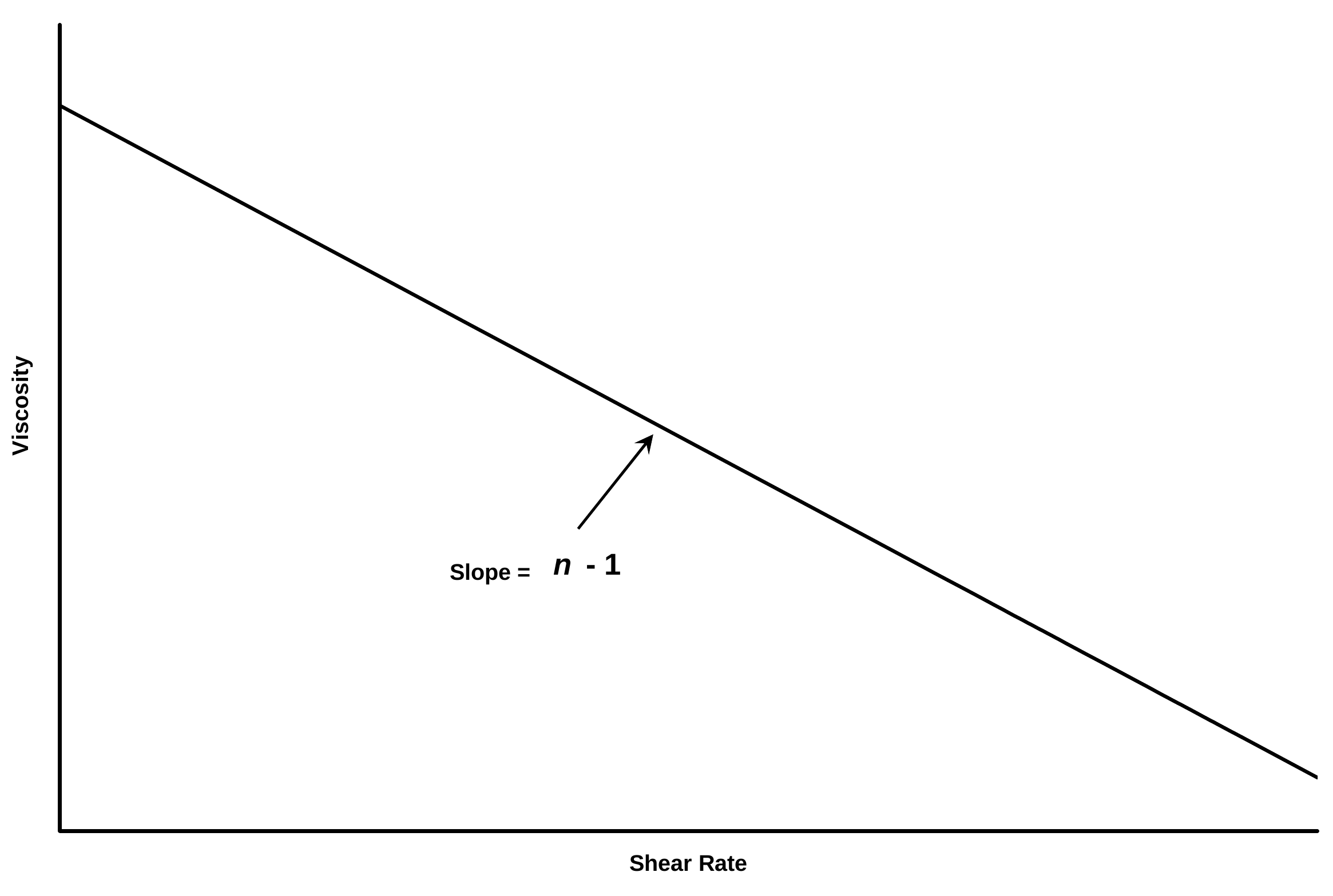}
  \caption[The bulk rheology of a \powlaw\ fluid on logarithmic scales for finite shear rates ($\sR>0$)]
  {The bulk rheology of a \powlaw\ fluid on logarithmic scales for finite shear rates ($\sR>0$).}
  \label{BulkRheoPowLaw}
\end{figure}

\subsubsection{\Ellis\ Model} \label{EllisSec}
This is a three-parameter model that describes \timeind\ \shThin\ non-\yields\ fluids. It is used
as a substitute for the \powlaw\ model and is appreciably better than the \powlaw\ in matching
experimental measurements. Its distinctive feature is the low-shear \NEW\ plateau without a
high-shear plateau. According to this model, the fluid viscosity $\Vis$ is given by \cite
{SadowskiB1965, Savins1969, BirdbookAH1987, CarreaubookKC1997}
\begin{equation} \label{EllisEq}
    \Vis = \frac{\lVis}{1+ \left(\frac{\sS}{\hsS} \right)^{\eAlpha - 1}}
\end{equation}
where $\lVis$ is the low-shear viscosity, $\sS$ is the shear stress, $\hsS$ is the shear stress at
which $\Vis=\lVis/2$ and $\eAlpha$ is an indicial parameter related to the \powlaw\ index by
$\eAlpha=1/n$. A generic graph demonstrating the bulk rheology, that is viscosity versus shear rate
on logarithmic scales, is shown in Figure (\ref{BulkRheoEllis}).


\begin{figure} [!h]
  \centering{}
  \includegraphics
  [scale=0.6]
  {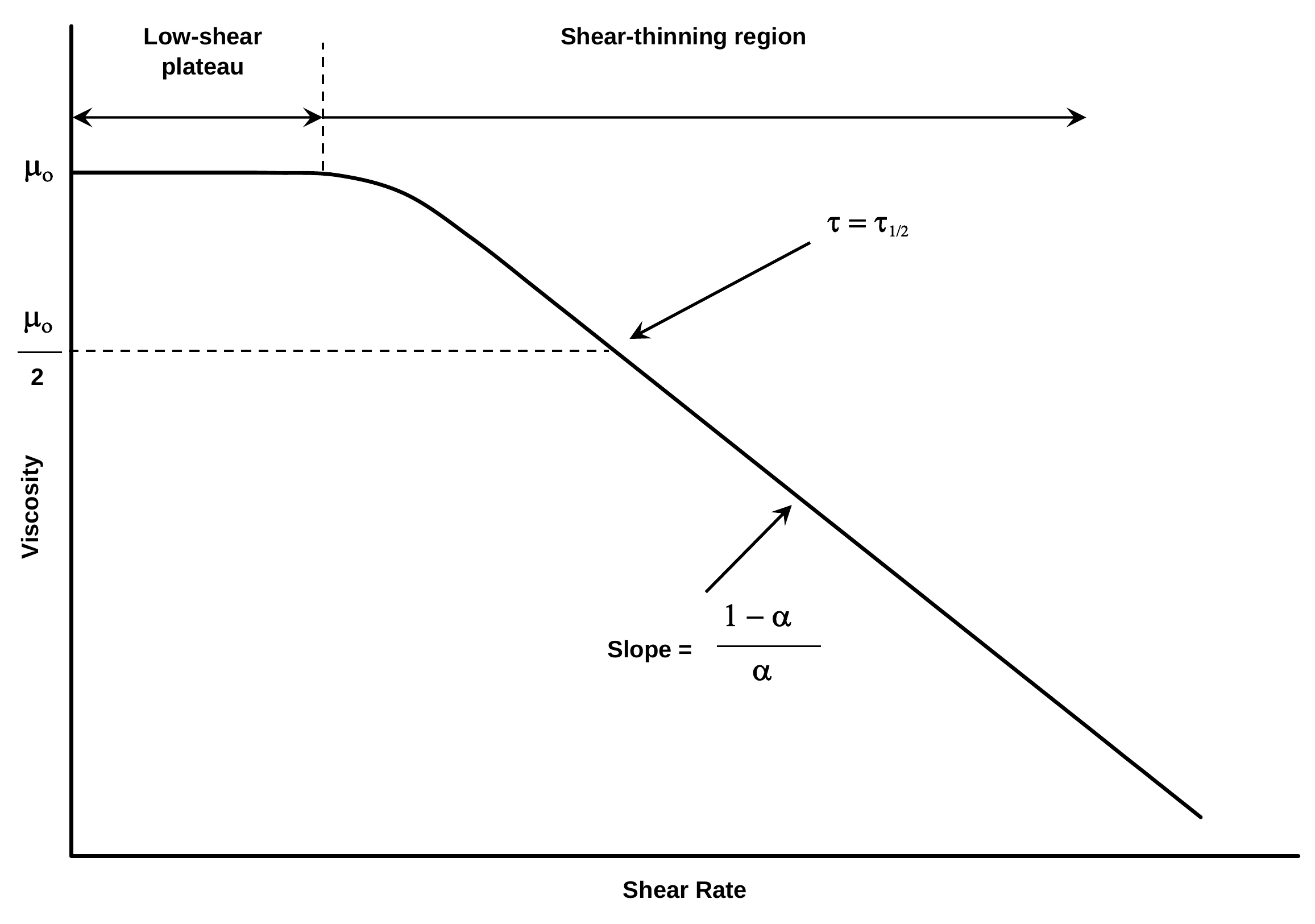}
  \caption[The bulk rheology of an \Ellis\ fluid on logarithmic scales for finite shear rates ($\sR>0$)]
  {The bulk rheology of an \Ellis\ fluid on logarithmic scales for finite shear rates ($\sR>0$).}
  \label{BulkRheoEllis}
\end{figure}

\subsubsection{\Carreau\ Model} \label{CarreauSec}
This is a four-parameter rheological model that can describe \shThin\ fluids with no \yields. It is
generally praised for its compliance with experiment. The distinctive feature of this model is the
presence of low- and high-shear plateaux. The \Carreau\ fluid is given by the relation
\cite{CarreaubookKC1997}
\begin{equation}\label{CarreauEq}
    \Vis = \hVis + \frac{\lVis - \hVis}
    {\left[1+(\sR t_{c})^{2}\right]^{\frac{1-n}{2}}}
\end{equation}
where $\Vis$ is the fluid viscosity, $\hVis$ is the viscosity at infinite shear rate, $\lVis$ is
the viscosity at zero shear rate, $\sR$ is the shear rate, $t_{c}$ is a characteristic time and $n$
is the flow behavior index. A generic graph demonstrating the bulk rheology is shown in Figure
(\ref{BulkRheoCarreau}).


\begin{figure} [!h]
  \centering{}
  \includegraphics
  [scale=0.6]
  {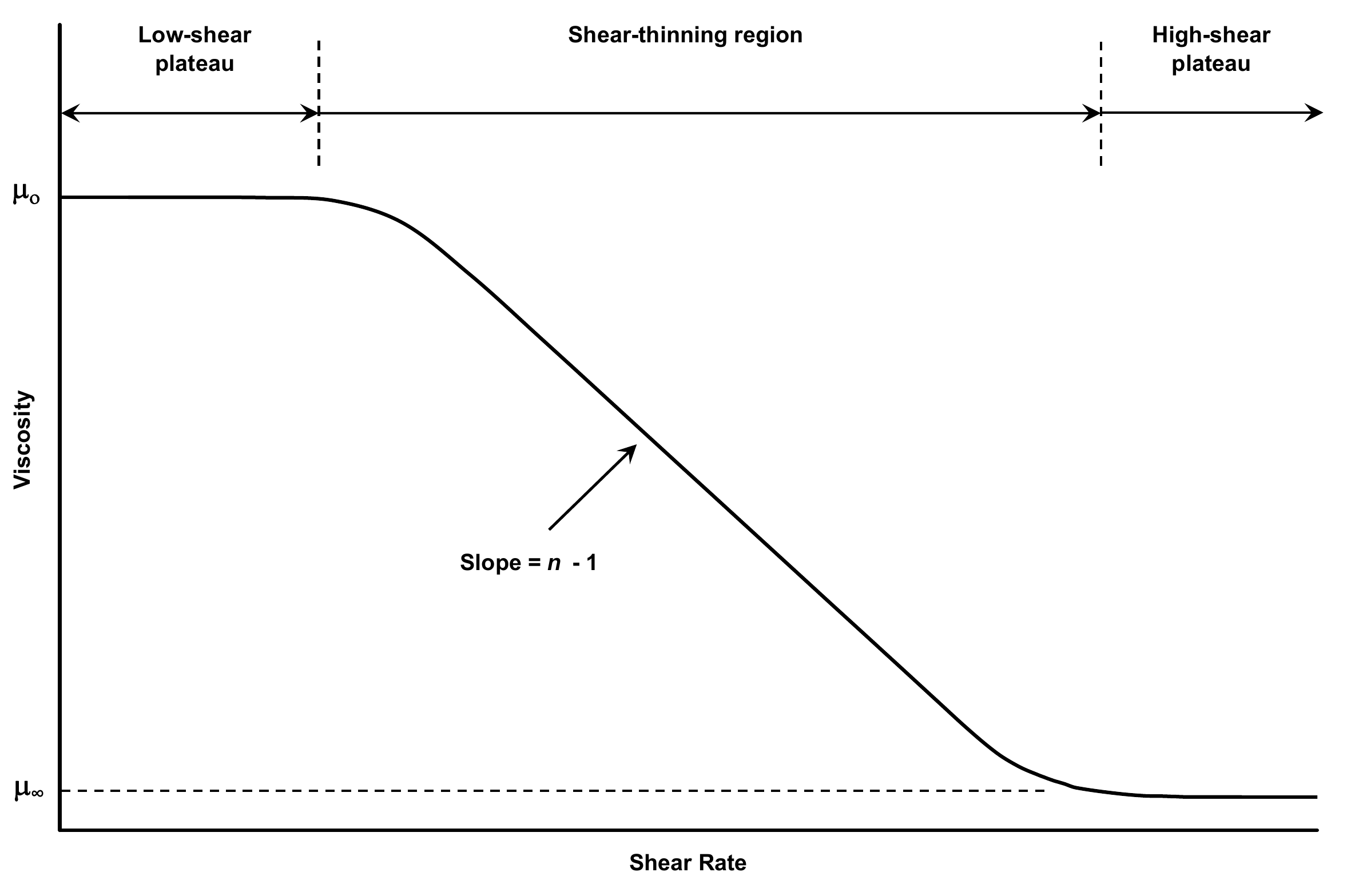}
  \caption[The bulk rheology of a \Carreau\ fluid on logarithmic scales for finite shear rates ($\sR>0$)]
  {The bulk rheology of a \Carreau\ fluid on logarithmic scales for finite shear rates ($\sR>0$).}
  \label{BulkRheoCarreau}
\end{figure}

\subsubsection{\HB\ Model} \label{HBSec}
The \HB\ is a simple rheological model with three parameters. Despite its simplicity it can
describe the \NEW\ and all main classes of the \timeind\ \nNEW\ fluids. It is given by the relation
\cite {Skellandbook1967, CarreaubookKC1997}
\begin{equation}\label{HBEq}
    \sS = \ysS + C \sR^{n}
    \verb|       | (\sS > \ysS)
\end{equation}
where $\sS$ is the shear stress, $\ysS$ is the \yields\ above which the substance starts to flow,
$C$ is the consistency factor, $\sR$ is the shear rate and $n$ is the flow behavior index. The \HB\
model reduces to the \powlaw\ when $\ysS=0$, to the \Bingham\ plastic when $n=1$, and to the
\Newton's law for viscous fluids when both these conditions are satisfied.

\vspace{0.2cm}

There are six main classes to this model:

\begin{enumerate}

\item \ShThin\ (\pseudoplastic)\hspace* {1.45cm} [$\ysS = 0, n < 1$]

\item \NEW\ \hspace* {4.9cm} [$\ysS = 0, n = 1$]

\item \ShThik\ (\dilatant) \hspace* {2.0cm} [$\ysS = 0, n > 1$]

\item \ShThin\ with \yields\ \hspace* {1.15cm} [$\ysS > 0, n < 1$]

\item \Bingham\ plastic \hspace* {3.95cm} [$\ysS > 0, n = 1$]

\item \ShThik\ with \yields\ \hspace* {0.8cm} [$\ysS > 0, n > 1$]

\end{enumerate}
These classes are graphically illustrated in Figure (\ref{TimeIndependent}).

\subsection{\Vc\ Fluids} \label{}
Polymeric fluids often show strong \vc\ effects. These include \shThin, \exThik, normal stresses,
and \timedep\ rheology. No theory is yet available that can adequately describe all of the observed
\vc\ phenomena in a variety of flows. Nonetheless, many differential and integral constitutive
models have been proposed in the literature to describe \vc\ flow. What is common to all these is
the presence of at least one characteristic time parameter to account for the fluid memory, that is
the stress at the present time depends upon the strain or rate of strain for all past times, but
with a fading memory \cite{Deiberthesis1978, Denn1990, Hulsenbook1996, OwensbookP2002,
Keunings2004}.

\vspace{0.2cm}

Broadly speaking, \vy\ is divided into two major fields: linear and nonlinear. {\bf Linear \vy} is
the field of rheology devoted to the study of \vc\ materials under very small strain or deformation
where the displacement gradients are very small and the flow regime can be described by a linear
relationship between stress and rate of strain. In principle, the strain has to be sufficiently
small so that the structure of the material remains unperturbed by the flow history. If the strain
rate is small enough, deviation from linear \vy\ may not occur at all. The equations of linear \vy\
are not valid for deformations of arbitrary magnitude and rate because they violate the principle
of frame invariance. The validity of linear \vy\ when the small-deformation condition is satisfied
with a large magnitude of rate of strain is still an open question, though it is widely accepted
that linear \vc\ constitutive equations are valid in general for any strain rate as long as the
total strain remains small. Nevertheless, the higher the strain rate the shorter the time at which
the critical strain for departure from linear regime is reached \cite{BirdbookAH1987,
CarreaubookKC1997, Larsonbook1999}.

\vspace{0.2cm}

The linear \vc\ models have several limitations. For example, they cannot describe strain rate
dependence of viscosity or normal stress phenomena since these are nonlinear effects. Due to the
restriction to infinitesimal deformations, the linear models may be more appropriate for the
description of \vc\ solids rather than \vc\ fluids. Despite the limitations of the linear \vc\
models and despite being of less interest to the study of flow where the material is usually
subject to large deformation, they are very important in the study of \vy\ for several reasons
\cite{BirdbookAH1987, Larsonbook1988, CarreaubookKC1997}:

\begin{enumerate}

\item They are used to characterize the behavior of \vc\ materials at
small deformations.

\item They serve as a motivation and starting point for developing
nonlinear models since the latter are generally extensions to the linears.

\item They are used for analyzing experimental data obtained in
small deformation experiments and for interpreting important \vc\ phenomena, at least
qualitatively.

\end{enumerate}

{\bf Non-linear \vy} is the field of rheology devoted to the study of \vc\ materials under large
deformation, and hence it is the subject of primary interest to the study of flow of \vc\ fluids.
Nonlinear \vc\ constitutive equations are sufficiently complex that very few flow problems can be
solved analytically. Moreover, there appears to be no differential or integral constitutive
equation general enough to explain the observed behavior of polymeric systems undergoing large
deformations but still simple enough to provide a basis for engineering design procedures
\cite{WhiteM1965, Skellandbook1967, BirdbookAH1987}.

\vspace{0.2cm}

As the linear \vy\ models are not valid for deformations of large magnitude because they do not
satisfy the principle of frame invariance, \Oldroyd\ and others tried to extend these models to
nonlinear regimes by developing a set of frame-invariant constitutive equations. These equations
define time derivatives in frames that deform with the material elements. Examples include
rotational, upper and lower convected time derivatives. The idea of these derivatives is to express
the constitutive equation in real space coordinates rather than local coordinates and hence
fulfilling the \Oldroyd's admissibility criteria for constitutive equations. These admissibility
criteria ensures that the equations are invariant under a change of coordinate system, value
invariant under a change of translational or rotational motion of the fluid element as it goes
through space, and value invariant under a change of rheological history of neighboring fluid
elements \cite{BirdbookAH1987, Larsonbook1988}.

\vspace{0.2cm}

There is a large number of rheological equations proposed for the description of nonlinear \vy, as
a quick survey to the literature reveals. However, many of these models are extensions or
modifications to others. The two most popular nonlinear \vc\ models in differential form are the
\UpCoMa\ and the \OldroydB\ models.

\vspace{0.2cm}

In the following sections we present two linear and three nonlinear \vc\ models in differential
form.

\subsubsection{Maxwell Model} \label{}
This is the first known attempt to obtain a \vc\ constitutive equation. This simple linear model,
with only one elastic parameter, combines the ideas of viscosity of fluids and elasticity of solids
to arrive at an equation for \vc\ materials \cite{BirdbookAH1987, MenaML1987}. \Maxwell\
\cite{Maxwell1867} proposed that fluids with both viscosity and elasticity can be described, in
modern notation, by the relation:
\begin{equation}\label{maxwell}
    {\sTen} + \rxTim \frac{\partial {\sTen}}{\partial t} =
    \lVis {\rsTen}
\end{equation}
where $\sTen$ is the extra stress tensor, $\rxTim$ is the fluid relaxation time, $t$ is time,
$\lVis$ is the low-shear viscosity and $\rsTen$ is the rate of strain tensor.

\subsubsection{\Jeffreys\ Model} \label{}
This is a linear model proposed as an extension to the \Maxwell\ model by including a time
derivative of the strain rate, that is \cite{Jeffreysbook1929, BirdbookAH1987}:
\begin{equation} \label{Jeffreys}
    {\sTen} + \rxTim \frac{\partial {\sTen}}{\partial t} =
    \lVis \left( {\rsTen} + \rdTim \frac{\partial {\rsTen}}{\partial t} \right)
\end{equation}
where $\rdTim$ is the retardation time that accounts for the corrections of this model and can be
seen as a measure of the time the material needs to respond to deformation. The \Jeffreys\ model
has three constants: a viscous parameter $\lVis$, and two elastic parameters,  $\rxTim$ and
$\rdTim$. The model reduces to the linear \Maxwell\ when $\rdTim = 0$, and to the \NEW\ when
$\rxTim = \rdTim = 0$. As observed by several authors, the \Jeffreys\ model is one of the most
suitable linear models to compare with experiment \cite{MardonesG1990}.

\subsubsection{\UpCoMa\ (UCM) Model} \label{}
To extend the linear \Maxwell\ model to the nonlinear regime, several time derivatives (e.g. upper
convected, lower convected and corotational) have been proposed to replace the ordinary time
derivative in the original model. The most commonly used of these derivatives in conjunction with
the \Maxwell\ model is the upper convected. On purely continuum mechanical grounds there is no
reason to prefer one of these \Maxwell\ equations to the others as they all satisfy frame
invariance. The popularity of the upper convected form is due to its more realistic features
\cite{Larsonbook1988, CarreaubookKC1997, OwensbookP2002}.

\vspace{0.2cm}

The \UpCoMa\ (UCM) is the simplest nonlinear \vc\ model and is one of the most popular models in
numerical modeling and simulation of \vc\ flow. Like its linear counterpart, it is a simple
combination of the \Newton's law for viscous fluids and the derivative of the \Hook's law for
elastic solids. Because of its simplicity, it does not fit the rich variety of \vc\ effects that
can be observed in complex rheological materials \cite{MardonesG1990}. However, it is largely used
as the basis for other more sophisticated \vc\ models. It represents, like the linear \Maxwell,
purely elastic fluids with shear-independent viscosity, i.e. \Boger\ fluids \cite{Tannerbook2000}.
UCM also predicts an elongation viscosity that is three times the shear viscosity, like \NEW, which
is unrealistic feature for most \vc\ fluids. The UCM model is obtained by replacing the partial
time derivative in the differential form of the linear \Maxwell\ with the upper convected time
derivative, that is
\begin{equation}\label{UCM1}
    {\sTen} + \rxTim {\ucd \sTen} = \lVis {\rsTen}
\end{equation}
where $\sTen$ is the extra stress tensor, $\rxTim$ is the relaxation time, $\lVis$ is the low-shear
viscosity, $\rsTen$ is the rate of strain tensor, and {$\ucd \sTen$} is the upper convected time
derivative of the stress tensor. This time derivative is given by
\begin{equation}\label{UCM2}
    \ucd \sTen =
    \frac{\partial {\sTen}}{\partial t} +
    \fVel \cdot \nabla \sTen -
    \left( \nabla \fVel \right)^{T} \cdot \sTen -
    \sTen \cdot \nabla \fVel
\end{equation}
where $t$ is time, $\fVel$ is the fluid velocity, $\left( \cdot \right)^{T}$ is the transpose of
tensor and $\nabla \fVel$ is the fluid velocity gradient tensor defined by Equation
(\ref{fVelGradTen}) in Appendix \ref{AppVE}. The convected derivative expresses the rate of change
as a fluid element moves and deforms. The first two terms in Equation (\ref{UCM2}) comprise the
material or substantial derivative of the extra stress tensor. This is the time derivative
following a material element and describes time changes taking place at a particular element of the
`material' or `substance'. The two other terms in (\ref{UCM2}) are the deformation terms. The
presence of these terms, which account for convection, rotation and stretching of the fluid motion,
ensures that the principle of frame invariance holds, that is the relationship between the stress
tensor and the deformation history does not depend on the particular coordinate system used for the
description \cite{BirdbookAH1987, CarreaubookKC1997, OsP2004}.

\vspace{0.2cm}

Despite the simplicity and limitations of the UCM model, it predicts important \vc\ properties such
as first normal stress difference in shear and strain hardening in elongation. It also predicts the
existence of stress relaxation after cessation of flow and elastic recoil. However, according to
the UCM the shear viscosity and the first normal stress difference are independent of shear rate
and hence the model fails to describe the behavior of most \vc\ fluids. Furthermore, it predicts a
\steadys\ elongational viscosity that becomes infinite at a finite elongation rate, which is
obviously far from physical reality \cite{Larsonbook1988}.

\subsubsection{\OldroydB\ Model} \label{OldroydB}
The \OldroydB\ model is a simple form of the more elaborate and rarely used \Oldroyd\ 8-constant
model which also contains the upper convected, the lower convected, and the corotational \Maxwell\
equations as special cases. \OldroydB\ is the second simplest nonlinear \vc\ model and is
apparently the most popular in \vc\ flow modeling and simulation. It is the nonlinear equivalent of
the linear \Jeffreys\ model, as it takes account of frame invariance in the nonlinear regime.
\OldroydB\ model can be obtained by replacing the partial time derivatives in the differential form
of the \Jeffreys\ model with the upper convected time derivatives \cite{BirdbookAH1987}
\begin{equation}\label{OBM1}
    {\sTen} + \rxTim {\ucd \sTen} =
    \lVis \left( {\rsTen} + \rdTim {\ucd \rsTen} \right)
\end{equation}
where {$\ucd \rsTen$} is the upper convected time derivative of the rate of strain tensor given by
\begin{equation}\label{OBM2}
    \ucd \rsTen =
    \frac{\partial {\rsTen}}{\partial t} +
    \fVel \cdot \nabla \rsTen -
    \left( \nabla \fVel \right)^{T} \cdot \rsTen -
    \rsTen \cdot \nabla \fVel
\end{equation}

\OldroydB\ model reduces to the \UCM\ when $\rdTim = 0$, and to \NEW\ when $\rxTim = \rdTim = 0$.
Despite the simplicity of the \OldroydB\ model, it shows good qualitative agreement with
experiments especially for dilute solutions of macromolecules and \Boger\ fluids. The model is able
to describe two of the main features of \vy, namely normal stress difference and stress relaxation.
It predicts a constant viscosity and first normal stress difference with zero second normal stress
difference. Moreover, the \OldroydB, like the UCM model, predicts an elongation viscosity that is
three times the shear viscosity, as it is the case in \NEW\ fluids. It also predicts an infinite
extensional viscosity at a finite extensional rate, which is physically unrealistic
\cite{BirdbookAH1987, MardonesG1990, Larsonbook1999, BalmforthbookC2001, OwensbookP2002}.

\vspace{0.2cm}

A major limitation of the UCM and \OldroydB\ models is that they do not allow for strain dependency
and second normal stress difference. To account for strain dependent viscosity and non-zero second
normal stress difference among other phenomena, more sophisticated models such as \Giesekus\ and
\PhanThienTanner\ (PTT) which introduce additional parameters should be considered. Such equations,
however, have rarely been used because of the theoretical and experimental complications they
introduce \cite{Boger1987}.

\subsubsection{\BauMan\ Model} \label{BautistaManero}
This model combines the \OldroydB\ constitutive equation for \vy\ (\ref{OBM1}) and the
\Fredrickson's kinetic equation for flow-induced structural changes which may by associated with
\thixotropy. The model requires six parameters that have physical significance and can be estimated
from rheological measurements. These parameters are the low and high shear rate viscosities, the
elastic modulus, the relaxation time, and two other constants describing the build up and break
down of viscosity.

\vspace{0.2cm}

The kinetic equation of \Fredrickson\ that accounts for the destruction and construction of
structure is given by
\begin{equation}\label{Fredrickson}
    \D \Vis t = \frac{\Vis}{\rxTimF} \left( 1 - \frac{\Vis}{\lVis} \right)
                + \kF \Vis \left( 1 - \frac{\Vis}{\hVis} \right) \sTen : \rsTen
\end{equation}
where  $\Vis$ is the viscosity, $t$ is the time of deformation, $\rxTimF$ is the relaxation time
upon the cessation of steady flow, $\lVis$ and $\hVis$ are the viscosities at zero and infinite
shear rates respectively, $\kF$ is a parameter that is related to a critical stress value below
which the material exhibits primary creep, $\sTen$ is the stress tensor and $\rsTen$ is the rate of
strain tensor. In this model $\rxTimF$ is a structural relaxation time whereas $\kF$ is a kinetic
constant for structure break down. The elastic modulus $\Go$ is related to these parameters by $\Go
= \Vis/\rxTimF$ \cite{BautistaSPM1999, BautistaSLPM2000, ManeroBSP2002, TardyA2005}.

\vspace{0.2cm}

The \BauMan\ model was originally proposed for the rheology of worm-like micellar solutions which
usually have an upper \NEW\ plateau and show strong signs of \shThin. The model, which incorporates
\shThin, elasticity and \thixotropy, can be used to describe the complex rheological behavior of
\vc\ systems that also exhibit \thixotropy\ and \rheopexy\ under shear flow. It predicts stress
relaxation, creep, and the presence of \thixotropic\ loops when the sample is subjected to \trans\
stress cycles. This model can also match steady shear, oscillatory and \trans\ measurements of \vc\
solutions \cite{BautistaSPM1999, BautistaSLPM2000, ManeroBSP2002, TardyA2005}.

\subsection{\TimeDep\ Fluids} \label{}
It is generally recognized that there are two main types of \timedep\ fluids: \thixotropic\ (work
softening) and \rheopectic\ (work hardening or anti-\thixotropic) depending upon whether the stress
decreases or increases with time at a given strain rate and constant temperature. There is also a
general consensus that the \timedep\ feature is caused by reversible structural change during the
flow process. However, there are many controversies about the details, and the theory of \timedep\
fluids is not well developed. Many models have been proposed in the literature to describe the
complex rheological behavior of \timedep\ fluids. Here we present only two models of this group.

\subsubsection{\Godfrey\ Model} \label{Godfrey}
\Godfrey\ \cite{Godfrey1973} suggested that at a particular shear rate the time dependence of
\thixotropic\ fluids can be described by the relation
\begin{equation}\label{godfrey}
    \Vis(t) = \iVis - \fdVis ( 1 - e^{-t/\fgTim} )
                    - \sdVis ( 1 - e^{-t/\sgTim} )
\end{equation}
where $\Vis(t)$ is the \timedep\ viscosity, $\iVis$ is the viscosity at the commencement of
deformation, $\fdVis$ and $\sdVis$ are the viscosity deficits associated with the decay time
constants $\fgTim$ and $\sgTim$ respectively, and $t$ is the time of shearing. The initial
viscosity specifies a maximum value while the viscosity deficits specify the reduction associated
with the particular time constants. As usual, the time constants define the time scales of the
process under examination. Although \Godfrey\ was originally proposed as a \thixotropic\ model it
can be easily extended to include \rheopexy.

\subsubsection{\SEM} \label{}
This is a general \timedep\ model that can describe both \thixotropic\ and \rheopectic\ rheologies
\cite{Barnes1997}. It is given by
\begin{equation}\label{SEM}
    \Vis(t) = \iVis + ( \inVis - \iVis ) ( 1 - e^{-(t/\seTim)^{c}} )
\end{equation}
where $\Vis(t)$ is the \timedep\ viscosity, $\iVis$ is the viscosity at the commencement of
deformation, $\inVis$ is the equilibrium viscosity at infinite time, $t$ is the time of
deformation, $\seTim$ is a time constant and $c$ is a dimensionless constant which in the simplest
case is unity.

\newpage

\section{Modeling the Flow of Fluids} \label{Modeling}
The basic equations describing the flow of fluids consist of the basic laws of continuum mechanics
which are the conservation principles of mass, energy and linear and angular momentum. These
governing equations indicate how the mass, energy and momentum of the fluid change with position
and time. The basic equations have to be supplemented by a suitable rheological equation of state,
or constitutive equation describing a particular fluid, which is a differential or integral
mathematical relationship that relates the extra stress tensor to the rate of strain tensor in
general flow condition and closes the set of governing equations. One then solves the constitutive
model together with the conservation laws using a suitable method to predict the velocity and
stress field of the flow \cite{Hulsenbook1986, BirdbookAH1987, Hulsenbook1990, CarreaubookKC1997,
Keunings2003, Keunings2004}.

\vspace{0.2cm}

In the case of \NS\ flows the constitutive equation is the \NEW\ stress relation \cite{OsP2004} as
given in (\ref{NewtonianTensForm}). In the case of more rheologically complex flows other \nNEW\
constitutive equations, such as \Ellis\ and \OldroydB, should be used to bridge the gap and obtain
the flow fields. To simplify the situation, several assumptions are usually made about the flow and
the fluid. Common assumptions include laminar, incompressible, \steadys\ and isothermal flow. The
last assumption, for instance, makes the energy conservation equation redundant.

\vspace{0.2cm}

The constitutive equation should be frame-invariant. Consequently, sophisticated mathematical
techniques are employed to satisfy this condition. No single choice of constitutive equation is
best for all purposes. A constitutive equation should be chosen considering several factors such as
the type of flow (shear or extension, steady or \trans, etc.), the important phenomena to capture,
the required level of accuracy, the available computational resources and so on. These
considerations can be influenced strongly by personal preference or bias. Ideally the rheological
equation of state is required to be as simple as possible, involving the minimum number of
variables and parameters, and yet having the capability to predict the behavior of complex fluids
in complex flows. So far, no constitutive equation has been developed that can adequately describe
the behavior of complex fluids in general flow situation \cite{Larsonbook1988, OwensbookP2002}.
Moreover, it is very unlikely that such an equation will be developed in the foreseeable future.

\subsection{Modeling Flow in Porous Media} \label{ModelingFlowPorous}
In the context of fluid flow, `porous medium' can be defined as a solid matrix through which small
interconnected cavities occupying a measurable fraction of its volume are distributed. These
cavities are of two types: large ones, called pores and throats, which contribute to the bulk flow
of fluid; and small ones, comparable to the size of the molecules, which do not have an impact on
the bulk flow though they may participate in other transportation phenomena like diffusion. The
complexities of the microscopic pore structure are usually avoided by resorting to macroscopic
physical properties to describe and characterize the porous medium. The macroscopic properties are
strongly related to the underlying microscopic structure. The best known examples of these
properties are the porosity and the permeability. The first describes the relative fractional
volume of the void space to the total volume while the second quantifies the capacity of the medium
to transmit fluid.

\vspace{0.2cm}

Another important property is the macroscopic homogeneity which may be defined as the absence of
local variation in the relevant macroscopic properties, such as permeability, on a scale comparable
to the size of the medium under consideration. Most natural and synthetic porous media have an
element of inhomogeneity as the structure of the porous medium is typically very complex with a
degree of randomness and can seldom be completely uniform. However, as long as the scale and
magnitude of these variations have a negligible impact on the macroscopic properties under
discussion, the medium can still be treated as homogeneous.

\vspace{0.2cm}

The mathematical description of the flow in porous media is extremely complex and involves many
approximations. So far, no analytical fluid mechanics solution to the flow through porous media has
been developed. Furthermore, such a solution is apparently out of reach in the foreseeable future.
Therefore, to investigate the flow through porous media other methodologies have been developed;
the main ones are the continuum approach, the bundle of tubes approach, numerical methods and
pore-scale network modeling. These approaches are outlined in the following sections with
particular emphasis on the flow of \nNEW\ fluids.

\subsubsection{Continuum Models} \label{}
These widely used models represent a simplified macroscopic approach in which the porous medium is
treated as a continuum. All the complexities and fine details of the microscopic pore structure are
absorbed into bulk terms like permeability that reflect average properties of the medium.
Semi-empirical equations such as \Darcy's law, \Blake-\Kozeny-\Carman\ or \Ergun\ equation fall
into this category. Commonly used forms of these equations are given in Table (\ref{continuumEqs}).
Several continuum models are based in their derivation on the capillary bundle concept which will
be discussed in the next section.


\begin{table} []
\centering %
\caption[Commonly used forms of the three popular continuum models] %
{Commonly used forms of the three popular continuum models.} %
\label{continuumEqs} %
\vspace{0.5cm} %
\begin{tabular}{ll}
\hline
{\bf Model \verb|                         |} & {\bf Equation} \\
\hline \vspace{-0.3cm} \\

    \Darcy                &  $\frac{\Delta P}{L} = \frac{\Vis q}{K}$ \\

   \Blake-\Kozeny-\Carman &  $\frac{\Delta P}{L} = \frac{72 C' \Vis q (1 - \epsilon)^{2}}{D_{p}^{2} \epsilon^{3}}$ \\

    \Ergun                &  $\frac{\Delta P}{L} = \frac{150 \Vis q}{D_{p}^{2}}   \frac{(1 - \epsilon)^{2}}{\epsilon^{3}} + \frac{1.75 \rho q^{2}}{D_{p}}   \frac{(1 - \epsilon)}{\epsilon^{3}}$ \\

\vspace{-0.4cm} \\ \hline
\end{tabular}
\end{table}


\vspace{0.2cm}

{\bf \Darcy's law} in its original form is a linear \NEW\ model that relates the local pressure
gradient in the flow direction to the fluid \supvel\ (i.e. volumetric flow rate per unit
cross-sectional area) through the viscosity of the fluid and the permeability of the medium.
\Darcy's law is apparently the simplest model for describing the flow in porous media, and is
widely used for this purpose. Although \Darcy's law is an empirical relation, it can be derived
from the capillary bundle model using \NS\ equation via homogenization (i.e. up-scaling microscopic
laws to a macroscopic scale). Different approaches have been proposed for the derivation of
\Darcy's law from first principles. However, theoretical analysis reveals that most of these rely
basically on the momentum balance equation of the fluid phase \cite{Schowalterbook1978}. \Darcy's
law is applicable to laminar flow at low \Reynolds\ number as it contains only viscous term with no
inertial term. Typically any flow with a \Reynolds\ number less than one is laminar. As the
velocity increases, the flow enters a nonlinear regime and the inertial effects are no longer
negligible. The linear \Darcy's law may also break down if the flow becomes vanishingly slow as
interaction between the fluid and the pore walls can become important. \Darcy\ flow model also
neglects the boundary effects and heat transfer. In fact the original \Darcy\ model neglects all
effects other than viscous \NEW\ effects. Therefore, the validity of \Darcy's law is restricted to
laminar, isothermal, purely viscous, incompressible \NEW\ flow. \Darcy's law has been modified
differently to accommodate more complex phenomena such as \nNEW\ and multi-phase flow. Various
generalizations to include nonlinearities like inertia have also been derived using homogenization
or volume averaging methods \cite{Shenoy1993}.

\vspace{0.4cm}

{\bf \Blake-\Kozeny-\Carman} (BKC) model for packed bed is one of the most popular models in the
field of fluid dynamics to describe the flow through porous media. It encompasses a number of
equations developed under various conditions and assumptions with obvious common features. This
family of equations correlates the pressure drop across a granular packed bed to the \supvel\ using
the fluid viscosity and the bed porosity and granule diameter. It is also used for modeling the
macroscopic properties of random porous media such as permeability. These semi-empirical relations
are based on the general framework of capillary bundle with various levels of sophistication. Some
forms in this family envisage the bed as a bundle of straight tubes of complicated cross-section
with an average hydraulic radius. Other forms depict the porous material as a bundle of tortuous
tangled capillary tubes for which the equation of \NS\ is applicable. The effect of tortuosity on
the average velocity in the flow channels gives more accurate portrayal of \nNEW\ flow in real beds
\cite{KozickiT1988, ChapuisA2003}. The BKC model is valid for laminar flow through packed beds at
low \Reynolds\ number where kinetic energy losses caused by frequent shifting of flow channels in
the bed are negligible. Empirical extension of this model to encompass transitional and turbulent
flow conditions has been reported in the literature \cite{ChhabraCM2001}.

\vspace{0.4cm}

{\bf Ergun} equation is a semi-empirical relation that links the pressure drop along a packed bed
to the \supvel. It is widely used to portray the flow through porous materials and to model their
physical properties. The input to the equation is the properties of the fluid (viscosity and mass
density) and the bed (length, porosity and granule diameter). Another, and very popular, form of
\Ergun\ equation correlates the \fricfact\ to the \Reynolds\ number. This form is widely used for
plotting experimental data and may be accused of disguising inaccuracies and defects. While \Darcy\
and \Blake-\Kozeny-\Carman\ contain only viscous term, the \Ergun\ equation contains both viscous
and inertial terms. The viscous term is dominant at low flow rates while the inertial term is
dominant at high flow rates \cite{Jones1976, Stevenson2003}. As a consequence of this duality,
\Ergun\ can reach flow regimes that are not accessible to \Darcy\ or BKC. A proposed derivation of
the \Ergun\ equation is based on a superposition of two asymptotic solutions, one for very low and
one for very high \Reynolds\ number flow. The lower limit, namely the BKC equation, is
quantitatively attributable to a fully developed laminar flow in a three-dimensional porous
structure, while in the higher limit the empirical Burke-Plummer equation for turbulent flow is
applied \cite{Plessis1994}.

\vspace{0.4cm}

The big advantage of the continuum approach is having a closed-form constitutive equation that
describes the highly complex phenomenon of flow through porous media using a few simple compact
terms. Consequently, the continuum models are easy to use with no computational cost. Nonetheless,
the continuum approach suffers from a major limitation by ignoring the physics of flow at pore
level.

\vspace{0.4cm}

Regarding \nNEW\ flow, most of these continuum models have been employed with some pertinent
modification to accommodate \nNEW\ behavior. A common approach for single-phase flow is to find a
suitable definition for the effective viscosity which will continue to have the dimensions and
physical significance of \NEW\ viscosity \cite{PearsonT2002}. However, many of these attempts,
theoretical and empirical, have enjoyed limited success in predicting the flow of
rheologically-complex fluids in porous media. Limitations of the \nNEW\ continuum models include
failure to incorporate \trans\ \timedep\ effects and to model \yields. Some of these issues will be
examined in the coming sections.

\vspace{0.4cm} 

The literature of \NEW\ and \nNEW\ continuum  models is overwhelming. In the following paragraph we
present a limited number of illuminating examples from the \nNEW\ literature.

\vspace{0.2cm}

\Darcy's law and \Ergun's equation have been used by Sadowski and Bird \cite{SadowskiB1965,
Sadowski1965, Sadowskithesis1963} in their theoretical and experimental investigations to \nNEW\
flow. They applied the \Ellis\ model to a \nNEW\ fluid flowing through a porous medium modeled by a
bundle of capillaries of varying cross-sections but of a definite length. This led to a generalized
form of \Darcy's law and of \Ergun's \fricfact\ correlation in each of which the \NEW\ viscosity
was replaced by an effective viscosity. A relaxation time term was also introduced as a correction
to account for \vc\ effects in the case of polymer solutions of high molecular weight. Gogarty
\etal\ \cite{GogartyLF1972} developed a modified form of the \nNEW\ \Darcy\ equation to predict the
viscous and elastic pressure drop versus flow rate relationships for the flow of elastic
\carboxymethylcellulose\ (CMC) solutions in beds of different geometry. Park \etal\
\cite{ParkHB1973, Parkthesis1972} used the \Ergun\ equation to correlate the pressure drop to the
flow rate for a \HB\ fluid flowing through packed beds by using an effective viscosity calculated
from the \HB\ model. To describe the non-steady flow of a \yields\ fluid in porous media, Pascal
\cite{Pascal1981} modified \Darcy's law by introducing a threshold pressure gradient to account for
\yields. This threshold gradient is directly proportional to the \yields\ and inversely
proportional to the square root of the absolute permeability. Al-Fariss and Pinder
\cite{AlfarissP1984, Alfariss1989} produced a general form of \Darcy's law by modifying the
\Blake-\Kozeny\ equation to describe the flow in porous media of \HB\ fluids. Chase and Dachavijit
\cite{ChaseD2003} modified the \Ergun\ equation to describe the flow of \yields\ fluids through
porous media using the bundle of capillary tubes approach.

\subsubsection{Capillary Bundle Models} \label{}
In the capillary bundle models the flow channels in a porous medium are depicted as a bundle of
tubes. The simplest form is the model with straight, cylindrical, identical parallel tubes oriented
in a single direction. \Darcy's law combined with the \Poiseuille's law give the following
relationship for the permeability of this model
\begin{equation}\label{BundleOfTubesKR}
    K = \frac{\epsilon R^{2}}{8}
\end{equation}
where $K$ and $\epsilon$ are the permeability and porosity of the bundle respectively, and $R$ is
the radius of the tubes.

\vspace{0.2cm}

The advantage of using this simple model, rather than other more sophisticated models, is its
simplicity and clarity. A limitation of the model is its disregard to the morphology of the pore
space and the heterogeneity of the medium. The model fails to reflect the highly complex structure
of the void space in real porous media. In fact the morphology of even the simplest medium cannot
be accurately depicted by this capillary model as the geometry and topology of real void space have
no similarity to a uniform bundle of tubes. The model also ignores the highly tortuous character of
the flow paths in real porous media with an important impact on the flow resistance and pressure
field. Tortuosity should also have consequences on the behavior of elastic and \yields\ fluids
\cite{KozickiT1988}. Moreover, as it is a unidirectional model its application is limited to simple
one-dimensional flow situations.

\vspace{0.2cm}

One important structural feature of real porous media that is not reflected in this model is the
\convdiv\ nature of the pore space which has a significant influence on the flow conduct of \vc\
and \yields\ fluids \cite{CannellaHS1988}. Although this simple model may be adequate for modeling
some cases of slow flow of purely viscous fluids through porous media, it does not allow the
prediction of an increase in the pressure drop when used with a \vc\ constitutive equation.
Presumably, the \convdiv\ nature of the flow field gives rise to an additional pressure drop, in
excess to that due to shearing forces, since porous media flow involves elongational flow
components. Therefore, a corrugated capillary bundle is a better choice for modeling such complex
flow fields \cite{PilitsisB1989, PearsonT2002}.

\vspace{0.2cm}

Ignoring the \convdiv\ nature of flow channels, among other geometrical and topological features,
can also be a source of failure for this model with respect to \yields\ fluids. As a straight
bundle of capillaries disregards the complex morphology of the pore space and because the yield
depends on the actual geometry and connectivity not on the flow conductance, the simple bundle
model will certainly fail to predict the yield point and flow rate of \yields\ fluids in porous
media. An obvious failure of this model is that it predicts a universal yield point at a particular
pressure drop, whereas in real porous media yield is a gradual process. Furthermore, possible bond
aggregation effects due to the connectivity and size distribution of the flow paths of the porous
media are completely ignored in this model.

\vspace{0.2cm}

Another limitation of this simple model is that the permeability is considered in the direction of
flow only, and hence may not correctly correspond to the permeability of the porous medium
especially when it is anisotropic. The model also fails to consider the pore size distributions.
Though this factor may not be of relevance in some cases where the absolute permeability is of
interest, it can be important in other cases such as \yields\ and certain phenomena associated with
two-phase flow \cite{Sorbiebook1991}.

\vspace{0.4cm}

To remedy the flaws of the uniform bundle of tubes model and to make it more realistic, several
elaborations have been suggested and used; these include

\begin{itemize}

\item Using a bundle of capillaries of varying cross-sections. This has
been employed by Sadowski and Bird \cite{SadowskiB1965} and other investigators. In fact the model
of corrugated capillaries is widely used especially for modeling \vc\ flow in porous media to
account for the excess pressure drop near the \convdiv\ regions. The \convdiv\ feature may also be
used when modeling the flow of \yields\ fluids in porous media.

\item Introducing an empirical tortuosity factor or using a tortuous
bundle of capillaries. For instance, in the \Blake-\Kozeny-\Carman\ model a tortuosity factor of
25/12 has been used and some forms of the BKC assumes a bundle of tangled capillaries
\cite{DudaHK1983, KozickiT1988}

\item To account for the 3D flow, the model can be improved by orienting
1/3 of the capillaries in each of the three spatial dimensions. The permeability, as given by
Equation (\ref{BundleOfTubesKR}), will therefore be reduced by a factor of 3 \cite{Sorbiebook1991}.

\item Subjecting the bundle radius size to a statistical distribution that
mimics the size distribution of the real porous media to be modeled by the bundle. Though we are
not aware of such a proposal in the literature, we believe it is sensible and viable.

\end{itemize}

It should be remarked that the simple capillary bundle model has been targeted by heavy criticism.
The model certainly suffers from several limitations and flaws as outlined earlier. However, it is
not different to other models and approaches used in modeling the flow through porous media as they
all are approximations with advantages and disadvantage, limitations and flaws. Like the rest, it
can be an adequate model when applied within its range of validity. Another remark is that the
capillary bundle model is better suited for describing porous media that are unconsolidated and
have high permeability \cite{ZaitounK1986}. This seems in line with the previous remark as the
capillary bundle is an appropriate model for this relatively simple porous medium with
comparatively plain structure.

\vspace{0.4cm}

Capillary bundle models have been widely used in the investigation of \NEW\ and \nNEW\ flow through
porous media with various levels of success. Past experience, however, reveals that no single
capillary model can be successfully applied in diverse structural and rheological situations. The
capillary bundle should therefore be designed and used with reference to the situation in hand.
Because the capillary bundle models are the basis for most continuum models, as the latter usually
rely in their derivation on some form of capillary models, most of the previous literature on the
use of the continuum approach applies to the capillary bundle. Here, we give a short list of some
contributions in this field in the context of \nNEW\ flow.

\vspace{0.2cm}

Sadowski \cite{Sadowski1965, Sadowskithesis1963} applied the \Ellis\ rheological model to a
corrugated capillary model of a porous medium to derive a modified \Darcy's law. Park
\cite{Parkthesis1972} used a capillary model in developing a generalized scale-up method to adapt
\Darcy's law to \nNEW\ fluids. Al-Fariss and Pinder \cite{AlfarissP1987} extended the capillary
bundle model to the \HB\ fluids and derived a form of the BKC model for \yields\ fluid flow through
porous media at low \Reynolds\ number. Vradis and Protopapas \cite{VradisP1993} extended a simple
capillary tubes model to describe the flow of \Bingham\ fluids in porous media and presented a
solution in which the flow is zero below a threshold head gradient and Darcian above it. Chase and
Dachavijit \cite{ChaseD2003} also applied the bundle of capillary tubes approach to model the flow
of a \yields\ fluid through a porous medium, similar to the approach of Al-Fariss and Pinder.

\subsubsection{Numerical Methods} \label{}
In the numerical approach, a detailed description of the porous medium at pore-scale with the
relevant physics at this level is used. To solve the flow field, numerical methods, such as finite
volume and finite difference, usually in conjunction with computational implementation are
utilized. The advantage of the numerical methods is that they are the most direct way to describe
the physical situation and the closest to full analytical solution. They are also capable, in
principle at least, to deal with \timedep\ \trans\ situations. The disadvantage is that they
require a detailed pore space description. Moreover, they are very complex and hard to implement
with a huge computational cost and serious convergence difficulties. Due to these complexities, the
flow processes and porous media that are currently within the reach of numerical investigation are
the most simple ones. These methods are also in use for investigating the flow in capillaries with
various geometries as a first step for modeling the flow in porous media and the effect of
corrugation on the flow field.

\vspace{0.4cm}

In the following we present a brief look with some examples from the literature for using numerical
methods to investigate the flow of \nNEW\ fluids in porous media. In many cases the corrugated
capillary was used to model the \convdiv\ nature of the passages in real porous media.

\vspace{0.2cm}

Talwar and Khomami \cite{TalwarK1992} developed a higher order Galerkin finite element scheme for
simulating a two-dimensional \vc\ flow and tested the algorithm on the flow of \UpCoMa\ (UCM) and
\OldroydB\ fluids in undulating tube. It was subsequently used to solve the problem of transverse
steady flow of these two fluids past an infinite square arrangement of infinitely long, rigid,
motionless cylinders. Souvaliotis and Beris \cite{SouvaliotisB1996} developed a domain
decomposition spectral collocation method for the solution of \steadys, nonlinear \vc\ flow
problems. The method was then applied to simulate \vc\ flows of UCM and \OldroydB\ fluids through
model porous media represented by a square array of cylinders and a single row of cylinders. Hua
and Schieber \cite{HuaS1998} used a combined finite element and Brownian dynamics technique
(CONNFFESSIT) to predict the \steadys\ \vc\ flow field around an infinite array of
squarely-arranged cylinders using two kinetic theory models. Fadili \etal\ \cite{FadiliTP2002}
derived an analytical 3D filtration formula for up-scaling isotropic \Darcy's flows of \powlaw\
fluids to heterogeneous \Darcy's flows by using stochastic homogenization techniques. They then
validated their formula by making a comparison with numerical experiments for 2D flows through a
rectangular porous medium using finite volume method.

\vspace{0.2cm}

Concerning the non-uniformity of flow channels and its impact on the flow which is an issue related
to the flow through porous media, numerical techniques have been used by a number of researchers to
investigate the flow of \vc\ fluids in \convdiv\ geometries. For example, Deiber and Schowalter
\cite{DeiberS1981, Deiberthesis1978} used the numerical technique of geometric iteration to solve
the nonlinear equations of motion for \vc\ flow through tubes with sinusoidal axial variations in
diameter. Pilitsis \etal\ \cite{PilitsisSB1991} carried a numerical simulation to the flow of a
\shThin\ \vc\ fluid through a periodically constricted tube using a variety of constitutive
rheological models. Momeni-Masuleh and Phillips \cite{MasulehP2004} used spectral methods to
investigate \vc\ flow in an undulating tube.

\subsubsection{Pore-Scale Network Modeling} \label{}
The field of network modeling is too diverse and complex to be fairly presented in this article. We
therefore present a general background on this subject followed by a rather detailed account of one
of the network models as an example. This is the model developed by Blunt and co-workers
\cite{Valvatnethesis2004, Lopezthesis2004, Sochithesis2007}. One reason for choosing this example
is because it is a well developed model that incorporates the main features of network modeling in
its current state. Moreover, it produced some of the best results in this field when compared to
the available theoretical models and experimental data. Because of the nature of this article, we
will focus on the single-phase \nNEW\ aspects.

\vspace{0.2cm}

Pore-scale network modeling is a relatively novel method that have been developed to deal with the
flow through porous media and other related issues. It can be seen as a compromise between the two
extremes of continuum and numerical approaches as it partly accounts for the physics of flow and
void space structure at pore level with affordable computational resources. Network modeling can be
used to describe a wide range of properties from capillary pressure characteristics to interfacial
area and mass transfer coefficients. The void space is described as a network of flow channels with
idealized geometry. Rules that determine the transport properties in these channels are
incorporated in the network to compute effective properties on a mesoscopic scale. The appropriate
pore-scale physics combined with a representative description of the pore space gives models that
can successfully predict average behavior \cite{Blunt2001, BluntJPV2002}.

\vspace{0.2cm}

Various network modeling methodologies have been developed and used. The general feature is the
representation of the pore space by a network of interconnected ducts (bonds or throats) of regular
shape and the use of a simplified form of the flow equations to describe the flow through the
network. A numerical solver is usually employed to solve a system of simultaneous equations to
determine the flow field. The network can be two-dimensional or three-dimensional with a random or
regular lattice structure such as cubic. The shape of the cylindrical ducts include circular,
square and triangular cross section (possibly with different shape factor) and may include
\convdiv\ feature. The network elements may contain, beside the conducting ducts, nodes (pores)
that can have zero or finite volume and may well serve a function in the flow phenomena or used as
junctions to connect the bonds. The simulated flow can be \NEW\ or \nNEW, single-phase, two-phase
or even three-phase. The physical properties of the flow and porous medium that can be obtained
from flow simulation include absolute and relative permeability, formation factor, resistivity
index, volumetric flow rate, apparent viscosity, threshold yield pressure and much more. Typical
size of the network is a few millimeters. One reason for this minute size is to reduce the
computational cost. Another reason is that this size is sufficient to represent a homogeneous
medium having an average throat size of the most common porous materials. Up-scaling the size of a
network is a trivial task if larger pore size is required. Moreover, extending the size of a
network model by attaching identical copies of the same model in any direction or imposing repeated
boundary conditions is another simple task.

\vspace{0.2cm}

The general strategy in network modeling is to use the bulk rheology of the fluid and the void
space description of the porous medium as an input to the model. The flow simulation in a network
starts by modeling the flow in a single capillary. The flow in a single capillary can be described
by the following general relation
\begin{equation}\label{generalFlowPresRel}
    Q = G' \Delta P
\end{equation}
where $Q$ is the volumetric flow rate, $G'$ is the flow conductance and $\Delta P$ is the pressure
drop. For a particular capillary and a specific fluid, $G'$ is given by
\begin{eqnarray}
  G' &=& G'(\mu) = \textrm{constant} \hspace{1.5cm}   \textrm{\NEW\ fluid} \nonumber \\
  G' &=& G'(\mu, \Delta P)           \hspace{2.75cm}  \textrm{Purely viscous \nNEW\ fluid} \nonumber \\
  G' &=& G'(\mu, \Delta P, t)        \hspace{2.45cm}  \textrm{Fluid with memory}
\end{eqnarray}

For a network of capillaries, a set of equations representing the capillaries and satisfying mass
conservation have to be solved simultaneously to find the pressure field and other physical
quantities. A numerical solver is usually employed in this process. For a network with $n$ nodes
there are $n$ equations in $n$ unknowns. These unknowns are the pressure values at the nodes. The
essence of these equations is the continuity of flow of incompressible fluid at each node in the
absence of sources and sinks. To find the pressure field, this set of equations have to be solved
subject to the boundary conditions which are the pressures at the inlet and outlet of the network.
This unique solution is `consistent' and `stable' as it is the only mathematically acceptable
solution to the problem, and, assuming the modeling process and the mathematical technicalities are
reliable, should mimic the unique physical reality of the pressure field in the porous medium.

\vspace{0.2cm}

For \NEW\ fluid, a single iteration is needed to solve the pressure field as the flow conductance
is known in advance because the viscosity is constant. For purely viscous \nNEW\ fluid, the process
starts with an initial guess for the viscosity, as it is unknown and pressure-dependent, followed
by solving the pressure field iteratively and updating the viscosity after each iteration cycle
until convergence is reached. For memory fluids, the dependence on time must be taken into account
when solving the pressure field iteratively. Apparently, there is no general strategy to deal with
this situation. However, for the \steadys\ flow of memory fluids of elastic nature a sensible
approach is to start with an initial guess for the flow rate and iterate, considering the effect of
the local pressure and viscosity variation due to \convdiv\ geometry, until convergence is
achieved. This approach was presented by Tardy and Anderson \cite{TardyA2005} and implemented with
some adaptation by Sochi \cite{Sochithesis2007}.

\vspace{0.2cm}

With regards to modeling the flow in porous media of complex fluids that have time dependency in a
dynamic sense due to \thixotropic\ or elastic nature, there are three major difficulties

\begin{itemize}

\item The difficulty of tracking the fluid elements in the pores and
throats and identifying their deformation history, as the path followed by these elements is random
and can have several unpredictable outcomes.

\item The mixing of fluid elements with various deformation history in the
individual pores and throats. As a result, the viscosity is not a well-defined property of the
fluid in the pores and throats.

\item The change of viscosity along the streamline since the deformation
history is continually changing over the path of each fluid element.

\end{itemize}
These complications can be ignored when dealing with cases of \steadys\ flow. Some suggestions on
network modeling of \timedep\ \thixotropic\ flow will be presented in Section (\ref{Thixotropy}).

\vspace{0.2cm}

Despite all these complications, network modeling in its current state is a simplistic approach
that models the flow in ideal situations. Most network models rely on various simplifying
assumptions and disregard important physical processes that have strong influence on the flow. One
such simplification is the use of geometrically uniform network elements to represent the flow
channels. \Vy\ and \yields\ are among the physical processes that are compromised by this
idealization of void space. Another simplification is the dissociation of the various flow
phenomena. For example, in modeling the flow of \yields\ fluids through porous media an implicit
assumption is usually made that there is no time dependency or \vy. This assumption, however, can
be reasonable in the case of modeling the dominant effect and may be valid in practical situations
where other effects are absent or insignificant. A third simplification is the adoption of
no-slip-at-wall condition. This widely accepted assumption means that the fluid at the boundary is
stagnant relative to the solid boundary. The effect of slip, which includes reducing shear-related
effects and influencing \yields\ behavior, is very important in certain circumstances and cannot be
ignored. However, this assumption is not far from reality in many other situations. Furthermore,
wall roughness, which is the rule in porous media, usually prevents wall slip or reduces its
effect. Other physical phenomena that can be incorporated in the network models to make it more
realistic include \retention\ (e.g. by \adsorption\ or \mectrap) and \wallexc. Most current network
models do not take account of these phenomena.

\vspace{0.4cm}

The model of Blunt and co-workers \cite{Valvatnethesis2004, Lopezthesis2004, Sochithesis2007},
which we present here as an example, uses three-dimensional networks built from a
topologically-equivalent three-dimensional voxel image of the pore space with the pore sizes,
shapes and connectivity reflecting the real medium. Pores and throats are modeled as having
triangular, square or circular cross-section by assigning a shape factor which is the ratio of the
area to the perimeter squared and obtained from the pore space image. Most of the network elements
are not circular. To account for the non-circularity when calculating the volumetric flow rate
analytically or numerically for a cylindrical capillary, an equivalent radius $R_{eq}$ is used
\begin{equation}
    \verb|       |    R_{eq} = \left( \frac{8G}{\pi} \right)^{1/4}
\end{equation}
where the geometric conductance, $G$, is obtained empirically from numerical simulation. The
network can be extracted from voxel images of a real porous material or from voxel images generated
by simulating the geological processes by which the porous medium was formed. Examples for the
latter are the two networks of Statoil which represent two different porous media: a \sandp\ and a
\Berea\ sandstone. These networks are constructed by {\O}ren and coworkers \cite{OrenBA1997,
OrenB2003} and are used by several researchers in flow simulation studies. Another possibility for
generating a network is by employing computational algorithms based on numeric statistical data
extracted from the porous medium of interest. Other possibilities can also be found in the
literature.

\vspace{0.2cm}

Assuming a laminar, isothermal and incompressible flow at low \Reynolds\ number, the only equations
that require attention are the constitutive equation for the particular fluid and the conservation
of volume as an expression for the conservation of mass. For \NEW\ flow, the pressure field can be
solved once and for all. For \nNEW\ flow, the situation is more complex as it involves
non-linearities and requires iterative techniques. For the simplest case of \timeind\ fluids, the
strategy is to start with an arbitrary guess. Because initially the pressure drop in each network
element is not known, an iterative method is used. This starts by assigning an effective viscosity
$\eVis$ to the fluid in each network element. The effective viscosity is defined as that viscosity
which makes \Poiseuille's equation fits any set of laminar flow conditions for \timeind\ fluids
\cite{Skellandbook1967}. By invoking the conservation of volume for incompressible fluid, the
pressure field across the entire network is solved using a numerical solver \cite{RugebookS1987}.
Knowing the pressure drops in the network, the effective viscosity of the fluid in each element is
updated using the expression for the flow rate with the \Poiseuille's law as a definition. The
pressure field is then recomputed using the updated viscosities and the iteration continues until
convergence is achieved when a specified error tolerance in the total flow rate between two
consecutive iteration cycles is reached. Finally, the total volumetric flow rate and the apparent
viscosity, defined as the viscosity calculated from the \Darcy's law, are obtained.

\vspace{0.2cm}

For the \steadys\ flow of \vc\ fluids, the approach of Tardy and Anderson \cite{TardyA2005,
Sochithesis2007} was used as outlined below

\begin{itemize}

\item The capillaries of the network are modeled with contraction to
account for the effect of \convdiv\ geometry on the flow. The reason is that the effects of fluid
memory take place on going through a radius change, as this change induces a change in strain rate
with a consequent viscosity changing effects. Examples of the \convdiv\ geometries are given in
Figure (\ref{ConvDivGeom}).

\item Each capillary is discretized in the flow direction and a
discretized form of the flow equations is used assuming a prior knowledge of the stress and
viscosity at the inlet of the network.

\item Starting with an initial guess for the flow rate and using iterative
technique, the pressure drop as a function of the flow rate is found for each capillary.

\item Finally, the pressure field for the whole network is found
iteratively until convergence is achieved. Once this happens, the flow rate through each capillary
in the network can be computed and the total flow rate through the network is determined by summing
and averaging the flow through the inlet and outlet capillaries.

\end{itemize}


\begin{figure}[!t]
  \centering{}
  \includegraphics
  [scale=0.35]
  {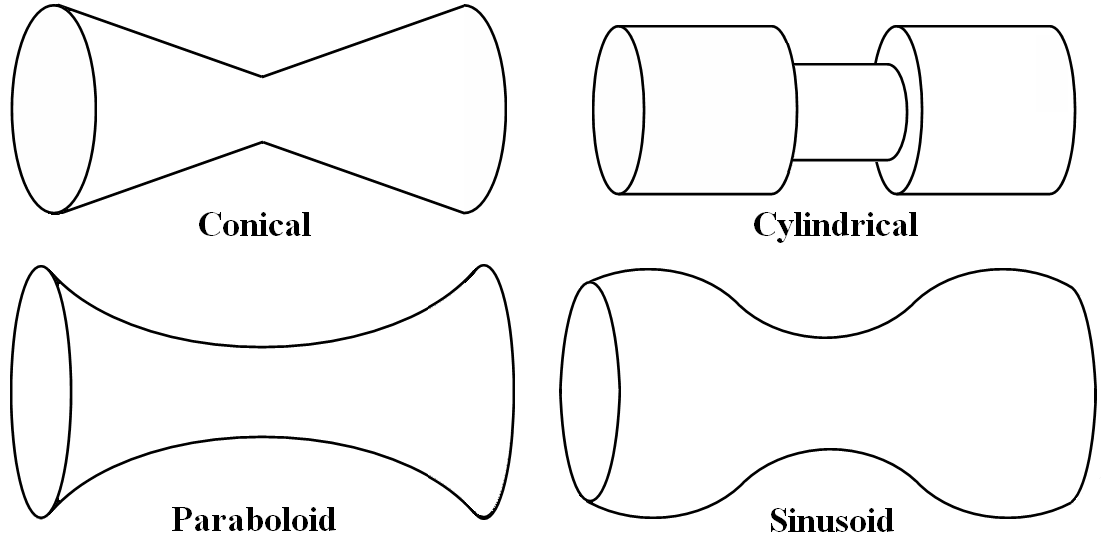}
  \caption[Examples of corrugated capillaries that can be used to model \convdiv\ geometry in porous media]
  {Examples of corrugated capillaries that can be used to model \convdiv\ geometry in porous media.}
  \label{ConvDivGeom}
\end{figure}


For \yields\ fluids, total blocking of the elements below their threshold yield pressure is not
allowed because if the pressure have to communicate, the substance before yield should be
considered a fluid with high viscosity. Therefore, to simulate the state of the fluid before yield
the viscosity is set to a very high but finite value so the flow virtually vanishes. As long as the
\yields\ substance is assumed to be fluid, the pressure field will be solved as for non-\yields\
fluids because the situation will not change fundamentally by the high viscosity. It is noteworthy
that the assumption of very high but finite zero-stress viscosity for \yields\ fluids is realistic
and supported by experimental evidence. A further condition is imposed before any element is
allowed to yield, that is the element must be part of a non-blocked path spanning the network from
the inlet to the outlet. What necessitates this condition is that any conducting element should
have a source on one side and a sink on the other, and this will not happen if either or both sides
are blocked.

\vspace{0.4cm}

Finally, a brief look at the literature of network modeling of \nNEW\ flow through porous media
will provide better insight into this field and its contribution to the subject. In their
investigation to the flow of \shThin\ fluids in porous media, Sorbie \etal\ \cite{SorbieCJ1989,
Sorbiebook1991} used 2D network models made up of connected arrays of capillaries to represent the
porous medium. The capillaries were placed on a regular rectangular lattice aligned with the
principal direction of flow, with constant bond lengths and coordination number 4, but the tube
radii are picked from a random distribution. The \nNEW\ flow in each network element was described
by the \Carreau\ model. Coombe \etal\ \cite{CoombeOB1993} used a network modeling approach to
analyze the impact of the \nNEW\ flow characteristics of polymers, foams, and emulsions at three
lengths and time scales. They employed a reservoir simulator to create networks of varying topology
in order to study the impact of porous media structure on rheological behavior. Shah \etal\
\cite{ShahKY1998} used small size 2D networks to study immiscible displacement in porous media
involving \powlaw\ fluids. Their pore network simulations include drainage where a \powlaw\ fluid
displaces a \NEW\ fluid at a constant flow rate, and the displacement of one \powlaw\ fluid by
another with the same \powlaw\ index. Tian and Yao \cite{TianY1999} used network models of
cylindrical capillaries on a 2D square lattice to study immiscible displacements of two-phase
\nNEW\ fluids in porous media. In this investigation the injected and the displaced fluids are
considered as \NEW\ and \nNEW\ \Bingham\ fluids respectively.

\vspace{0.2cm}

To study the flow of \powlaw\ fluids through porous media, Pearson and Tardy \cite{PearsonT2002}
employed a 2D square-grid network of capillaries, where the radius of each capillary is randomly
distributed according to a predefined probability distribution function. Their investigation
included the effect of the rheological and structural parameters on the \nNEW\ flow behavior.
Tsakiroglou \cite{Tsakiroglou2002, Tsakiroglou2004} and Tsakiroglou \etal\
\cite{TsakiroglouTKPS2003} used pore network models on 2D square lattices to investigate various
aspects of single- and two-phase flow of \nNEW\ fluids in porous media. The rheological models
which were employed in this investigation include \Meter, \powlaw\ and mixed \Meter\ and \powlaw.
Lopez \etal\ \cite{LopezVB2003, LopezB2004, Lopezthesis2004} investigated single- and two-phase
flow of \shThin\ fluids in porous media using network modeling. They used morphologically complex
3D networks to model the porous media, while the \shThin\ rheology was described by \Carreau\
model. Balhoff and Thompson \cite{BalhoffT2004, BalhoffT2006, Balhoffthesis2005} employed 3D random
network models extracted from computer-generated random sphere packing to investigate the flow of
\shThin\ and \yields\ fluids in packed beds. The \powlaw, \Ellis\ and \HB\ fluids were used as
rheological models. To model \nNEW\ flow in the throats, they used analytical expressions for a
capillary tube with empirical tuning to key parameters to represent the throat geometry and
simulate the fluid dynamics. Perrin \etal\ \cite{PerrinTSC2006} used network modeling with
\Carreau\ rheology to analyze their experimental findings on the flow of hydrolyzed
\polyacrylamide\ solutions in etched silicon micromodels. The network simulation was performed on a
2D network model with a range of rectangular channel widths exactly as in the physical micromodel
experiment. Sochi and Blunt \cite{SochiB2008, Sochithesis2007} used network modeling to study
single-phase \nNEW\ flow in porous media. Their investigation includes modeling the \Ellis\ and
\HB\ as \timeind\ rheological models and the \BauMan\ as a \vc\ model with \thixotropic\ features.
The \nNEW\ phenomena that can be described by these models include \shThin, \shThik, \yields,
\thixotropy\ and \vy.

\vspace{0.2cm}

Network modeling was also used by several researchers to investigate the generation and
mobilization of foams and \yields\ fluids in porous media. These include Rossen and Mamun
\cite{RossenM1993} and Chen \etal\ \cite{ChenYR2004, ChenYR2005, ChenRY2005}.

\newpage

\section{\YieldS} \label{Yield}
\Yields\ or \viscoplastic\ fluids are characterized by their ability to sustain shear stresses,
that is a certain amount of stress must be exceeded before the flow initiates. So an ideal \yields\
fluid is a solid before yield and a fluid after. Accordingly, the viscosity of the substance
changes from an infinite to a finite value. However, the physical situation indicates that it is
more realistic to regard a \yields\ substance as a fluid whose viscosity as a function of applied
stress has a discontinuity as it drops sharply from a very high value on exceeding a critical
stress.

\vspace{0.2cm}

There are many controversies and unsettled issues in the \nNEW\ literature about \yields\
phenomenon and \yields\ fluids. In fact, even the concept of a \yields\ has received much recent
criticism, with evidence presented to suggest that most materials weakly yield or creep near zero
strain rate. The supporting argument is that any material will flow provided that one waits long
enough. These conceptual difficulties are backed by practical and experimental complications. For
example, the value of \yields\ for a particular fluid is difficult to measure consistently and it
may vary by more than an order of magnitude depending on the measurement technique
\cite{BirdbookAH1987, CarreaubookKC1997, Barnes1999, BalmforthbookC2001}. Several constitutive
equations to describe liquids with \yields\ are in use; the most popular ones are \Bingham,
\Casson\ and \HB. Some have suggested that the \yields\ values obtained via such models should be
considered model parameters and not real material properties.

\vspace{0.2cm}

There are several difficulties in working with \yields\ fluids and validating the experimental
data. One difficulty is that \yields\ value is usually obtained by extrapolating a plot of shear
stress to zero shear rate \cite{ParkHB1973, AlfarissP1984, CarreaubookKC1997}. This can result in a
variety of values for \yields, depending on the distance from the shear stress axis experimentally
accessible by the instrument used. The vast majority of \yields\ data reported results from such
extrapolations, making most values in the literature instrument-dependent \cite{CarreaubookKC1997}.
Another method used to measure \yields\ is by lowering the shear rate until the shear stress
approaches a constant. This may be described as the dynamic \yields\ \cite{Larsonbook1999}. The
results obtained using such methods may not agree with the static \yields\ measured directly
without disturbing the microstructure during the measurement. The latter seems more relevant to the
flow initiation under gradual increase in pressure gradient as in the case of flow in porous media.
Consequently, the accuracy of the predictions made using flow simulation models in conjunction with
such experimental data is limited.

\vspace{0.2cm}

Another difficulty is that while in the case of pipe flow the \yields\ value is a property of the
fluid, in the case of flow in porous media there are strong indications that in a number of
situations it may depend on both the fluid and the porous medium itself \cite{Bearbook1972,
VradisP1993}. One possible explanation is that \yields\ value may depend on the size and shape of
the pore space when the polymer molecules become comparable in size to the pore. The implicit
assumption that \yields\ value at pore-scale is the same as the value at bulk may not be self
evident. This makes the predictions of the models based on analytical solution to the flow in a
uniformly-shaped tube combined with the bulk rheology less accurate. When the duct size is small,
as it is usually the case in porous media, flow of macromolecule solutions normally displays
deviations from predictions based on corresponding viscometric data \cite{LiuM1998}. Moreover, the
highly complex shape of flow paths in porous media must have a strong impact on the actual yield
point, and this feature is lost by modeling these paths with ducts of idealized geometry.
Consequently, the concept of equivalent radius $R_{eq}$, which is used in network modeling, though
is completely appropriate for \NEW\ fluids and reasonably appropriate for purely viscous \nNEW\
fluids with no \yields, seems inappropriate for \yields\ fluids as yield depends on the actual
shape of the void space rather than the equivalent radius and flow conductance.

\subsection{Modeling \YieldS\ in Porous Media} \label{}
In this section we outline the failure of the four approaches for modeling the flow through porous
media in dealing with the flow of \yields\ fluids. It is apparent that no continuum model can
predict the yield point of a \yields\ fluid in complex porous media. The reason is that these
models do not account for the complex geometry and topology of the void space. As the yield point
depends on the fine details of the pore space structure, no continuum model is expected to predict
the threshold yield pressure (TYP) of \yields\ fluids in porous media. The continuum models also
fail to predict the flow rate, at least at transition stage where the medium is partly conducting,
because according to these models the medium is either fully blocked or fully flowing whereas in
reality the yield of porous medium is a gradual process.

\vspace{0.2cm}

Regarding the capillary bundle models, the situation is similar to the continuum models as they
predict a single universal yield if a uniform bundle of capillaries is assumed. Moreover, because
all capillary bundle models fail to capture the topology and geometry of complex porous media they
cannot predict the yield point and describe the flow rate of \yields\ fluids in porous media since
yield is highly dependent on the fine details of the void space. An important aspect of the
geometry of real porous media which strongly affects the yield point and flow rate of \yields\
fluids is the \convdiv\ nature of the flow paths. This feature is not reflected by the bundle of
uniform capillaries models. Another feature is the connectivity of the flow channels where bond
aggregation (i.e. how the throats are distributed and arranged) strongly affects yield behavior.

\vspace{0.2cm}

Concerning the application of numerical methods to \yields\ fluids in porous media, very few
studies can be found on this subject (e.g. \cite{Vikhansky2007}). Moreover, the results, which are
reported only for very simple cases of porous media, cannot be fully assessed. It seems therefore
that network modeling is the only viable candidate for modeling \yields\ fluids in porous media.
However, because the research in this field is limited, no final conclusion on the merit of this
approach can be reached. Nonetheless, the modest success in modeling \yields\ as experienced by
some investigators (e.g. Balhoff \cite{Balhoffthesis2005} and Sochi \cite{Sochithesis2007})
indicates that network modeling in its current state is not capable of dealing with such a complex
phenomenon. One possible reason is the comparative simplicity of the rheological models, such as
\Bingham, used in these investigations. These models can possibly offer a phenomenological
description of \yields\ in simple flow situations but are certainly unable to accommodate the
underlying physics at pore level in complex porous media. Consequently, \yields\ as a model
parameter obtained in bulk viscometry may not be appropriate to use in this complex situation.
Another reason is the experimental difficulties associated with \yields\ fluids. This can make the
experimental results subject to complications, such as \vy\ and \retention, that may not be
accounted for in the network model. A third reason is the relative simplicity of the current
network modeling approach to \yields\ fluids. This is supported by the fact that better results are
usually obtained for non-\yields\ fluids using the same network modeling techniques. One major
limitation of the current network models with regard to \yields\ fluids is the use of analytical
expressions for cylindrical tubes based on the concept of equivalent radius $R_{eq}$. This is far
from reality where the void space retains highly complex shape and connectivity. Consequently, the
yield condition for cylindrical capillaries becomes invalid approximation to the yield condition
for the intricate flow paths in porous media.

\vspace{0.2cm}

In summary, \yields\ fluid results are extremely sensitive to how the fluid is characterized, how
the void space is described and how the yield process is modeled. In the absence of a comprehensive
and precise incorporation of all these factors in the network model, pore-scale modeling of
\yields\ fluids in porous media remains a crude approximation that may not produce quantitatively
sensible predictions. The final conclusion is that \yields\ is a problematic phenomenon, and hence
very modest success has been achieved in this area by any one of the four modeling approaches.

\subsection{Predicting Threshold Yield Pressure (TYP)} \label{}
Here we discuss the attempts to predict the yield point of a complex porous medium from the void
space description and \yields\ value of an ideal \yields\ fluid without modeling the flow process.
In the literature of \yields\ we can find two well-developed methods proposed for predicting the
yield point of a morphologically-complex network that depicts a porous medium; these are the
Minimum Threshold Path (MTP) and the \percolation\ theory. In this regard, there is an implicit
assumption that the network is an exact replica of the medium and the \yields\ value reflects the
\yields\ of real fluid so that any failure of these algorithms can not be attributed to mismatch or
any factor other than flaws in these algorithms. It should be remarked that the validity of these
methods can be tested by experiment.

\vspace{0.2cm}

Predicting the threshold yield pressure of a \yields\ fluid in porous media in its simplest form
may be regarded as a special case of the more general problem of finding the threshold conducting
path in disordered media that consist of elements with randomly distributed thresholds. This
problem was analyzed by Roux and Hansen \cite {RouxH1987} in the context of studying the conduction
of an electric network of diodes by considering two different cases, one in which the path is
directed (no backtracking) and one in which it is not. They suggested that the minimum overall
threshold potential difference across the network is akin to a \percolation\ threshold and
investigated its dependence on the lattice size. Kharabaf and Yortsos \cite {KharabafY1997} noticed
that a firm connection of the lattice-threshold problem to \percolation\ appears to be lacking and
the relation of the Minimum Threshold Path (MTP) to the minimum path of \percolation, if it indeed
exists, is not self-evident. They presented a new algorithm, \InvPM\ (IPM), for the construction of
the MTP, based on which its properties can be studied.  The \InvPM\ method was further extended by
Chen \etal\ \cite {ChenRY2005} to incorporate dynamic effects due to viscous friction following the
onset of mobilization.

\vspace{0.2cm}

The IPM is an algorithm for finding the inlet-to-outlet path that minimizes the sum of values of a
property assigned to the individual elements of the network, and hence finding this minimum. For a
\yields\ fluid, this reduces to finding the inlet-to-outlet path that minimizes the yield pressure.
The yield pressure of this path is then taken as the network threshold yield pressure. A detailed
description of this algorithm is given in \cite{KharabafY1997, Sochithesis2007}. Other algorithms
that achieve this minimization process can be found in the literature. However, they all rely on a
similar assumption to that upon which the IPM is based, that is the threshold yield pressure of a
network is the minimum sum of the threshold yield pressures of the individual elements of all
possible paths from the inlet to the outlet.

\vspace{0.2cm}

There are two possibilities for defining yield stress fluids before yield: either solid-like
substances or liquids with very high viscosity. According to the first, the most sensible way for
modeling a presumed pressure gradient inside a medium is to be a constant, that is the pressure
drop across the medium is a linear function of the spatial coordinate in the flow direction, as any
other assumption needs justification. Whereas in the second case the fluid should be treated like
non-\yields\ fluids and hence the pressure field inside the porous medium should be subject to the
consistency criterion for the pressure field which was introduced earlier. The logic is that the
magnitude of the viscosity should have no effect on the flow conduct as long as the material is
assumed to be fluid.

\vspace{0.2cm}

Several arguments can be presented against the MTP algorithms for predicting the yield point of a
medium or a network. Though certain arguments may be more obvious for a network with cylindrical
ducts, they are valid in general for regular and irregular geometries of flow channels. Some of
these arguments are outlined below

\begin{itemize}

\item The MTP algorithms are based on the assumption that the threshold
yield pressure (TYP) of an ensemble of serially connected bonds is the sum of their yield
pressures. This assumption can be challenged by the fact that for a non-uniform ensemble (i.e. an
ensemble whose elements have different TYPs) the pressure gradient across the ensemble should reach
the threshold yield gradient of the bottleneck (i.e. the element with the highest TYP) of the
ensemble if yield should occur. Consequently, the TYP of the ensemble will be higher than the sum
of the TYPs of the individual elements. This argument may be more obvious if \yields\ fluids are
regarded as solids and a linear pressure drop is assumed.

\item Assuming the \yields\ substances before yield are fluids with
high viscosity, the dynamic aspects of the pressure field are neglected because the aim of the MTP
algorithms is to find a collection of bonds inside the network with a certain condition based on
the intrinsic properties of these elements irrespective of the pressure field. The reality is that
the bonds are part of a network that is subject to a pressure field, so the pressure across each
individual element must comply with a dynamically stable pressure configuration over the whole
network. The MTP algorithms rely for justification on a hidden assumption that the minimum sum
condition is identical or equivalent to the stable configuration condition, which is not obvious,
because it is very unlikely that a stable configuration will require a pressure drop across each
element of the path of minimum sum that is identical to the threshold yield pressure of that
element. Therefor, it is not clear that the actual path of yield must coincide, totally or
partially, with the path of the MTP algorithms let alone that the actual value of yield pressure
must be predicted by these methods. To sum up, these algorithms disregard the global pressure field
which can communicate with the internal nodes of the serially connected ensemble as it is part of a
network and not an isolated collection of bonds. It is not obvious that the condition of these
algorithms should agree with the requirement of a stable and mathematically consistent pressure
filed as defined in Section (\ref{ModelingFlowPorous}). Such an agreement should be regarded as an
extremely unlikely coincidence.

\item The MTP algorithms that allows backtracking have another
drawback that is in some cases the path of minimum sum requires a physically unacceptable pressure
configuration. This may be more obvious if the \yields\ substances are assumed to be solid before
yield though it is still valid with the fluid assumption.

\item The effect of tortuosity is ignored since it is implicitly assumed
that the path of yield is an ensemble of serially-connected and linearly-aligned tubes, whereas in
reality the path is generally tortuous as it is part of a network and can communicate with the
global pressure field through the intermediate nodes. The effect of tortuosity, which is more
obvious for the solid assumption, is a possible increase in the external threshold pressure
gradient and a possible change in the bottleneck.

\end{itemize}

\vspace{0.4cm}

Concerning the \percolation\ approach, it is tempting to consider the conduct of \yields\ fluids in
porous media as a \percolation\ phenomenon to be analyzed by the classical \percolation\ theory.
However, three reasons, at least, may suggest otherwise

\begin{enumerate}

\item The conventional \percolation\ models can be applied only if the
conducting elements are homogeneous, i.e. it is assumed in these models that the intrinsic property
of all elements in the network are equal. However, this assumption cannot be justified for most
kinds of media where the elements vary in their conduction and yield properties. Therefore, to
apply \percolation\ principles, a generalization to the conventional \percolation\ theory is needed
as suggested by Selyakov and Kadet \cite {SelyakovbookK1996}.

\item The network elements cannot yield independently as a spanning path
bridging the inlet to the outlet is a necessary condition for yield. This contradicts the
\percolation\ theory assumption that the elements conduct independently.

\item The pure \percolation\ approach ignores the dynamic aspects of the
pressure field, that is a stable pressure configuration is a necessary condition which may not
coincide with the simple \percolation\ requirement. The \percolation\ condition, as required by the
classic \percolation\ theory, determines the stage at which the network starts flowing according to
the intrinsic properties of its elements as an ensemble of conducting bonds regardless of the
dynamic aspects introduced by the external pressure field. This argument is more obvious if
\yields\ substances are regarded as fluids with high viscosity.

\end{enumerate}

In a series of studies on generation and mobilization of foam in porous media, Rossen \etal\  \cite
{RossenG1990, RossenM1993} analyzed the threshold yield pressure using \percolation\ theory
concepts and suggested a simple \percolation\ model. In this model, the \percolation\ cluster is
first found, then the MTP was approximated as a subset of this cluster that samples those bonds
with the smallest individual thresholds \cite {ChenRY2005}. This approach relies on the validity of
applying \percolation\ theory to \yields, which is disputed. Moreover, it is a mere coincidence if
the yield path is contained within the \percolation\ sample. Yield is an on/off process which
critically depends on factors other than smallness of individual thresholds. These factors include
the particular distribution and configuration of these elements, being within a larger network and
hence being able to communicate with the global pressure field, and the dynamic aspects of the
pressure field and stability requirement. Any approximation, therefore, has very little meaning in
this context.

\newpage

\section{\Vy} \label{Viscoelasticity}
\Vc\ substances exhibit a dual nature of behavior by showing signs of both viscous fluids and
elastic solids. In its most simple form, \vy\ can be modeled by combining \Newton's law for viscous
fluids (stress $\propto$ rate of strain) with \Hook's law for elastic solids (stress $\propto$
strain), as given by the original \Maxwell\ model and extended by the Convected \Maxwell\ models
for the nonlinear \vc\ fluids. Although this idealization predicts several basic \vc\ phenomena, it
does so only qualitatively. The behavior of \vc\ fluids is drastically different from that of \NEW\
and inelastic \nNEW\ fluids. This includes the presence of normal stresses in shear flows,
sensitivity to deformation type, and memory effects such as stress relaxation and \timedep\
viscosity. These features underlie the observed peculiar \vc\ phenomena such as rod-climbing
(\Weissenberg\ effect), die swell and open-channel siphon \cite{Larsonbook1988}. Most \vc\ fluids
exhibit \shThin\ and an elongational viscosity that is both strain and extensional strain rate
dependent, in contrast to \NEW\ fluids where the elongational viscosity is constant and in
proportion to shear viscosity \cite{Boger1987}.

\vspace{0.2cm}

The behavior of \vc\ fluids at any time is dependent on their recent deformation history, that is
they possess a fading memory of their past. Indeed a material that has no memory cannot be elastic,
since it has no way of remembering its original shape. Consequently, an ideal \vc\ fluid should
behave as an elastic solid in sufficiently rapid deformations and as a \NEW\ liquid in sufficiently
slow deformations. The justification is that the larger the strain rate, the more strain is imposed
on the sample within the memory span of the fluid \cite{Boger1987, BirdbookAH1987, Larsonbook1988}.
Many materials are \vc\ but at different time scales that may not be reached. Dependent on the time
scale of the flow, \vc\ materials show mainly viscous or elastic behavior. The particular response
of a sample in a given experiment depends on the time scale of the experiment in relation to a
natural time of the material. Thus, if the experiment is relatively slow, the sample will appear to
be viscous rather than elastic, whereas, if the experiment is relatively fast, it will appear to be
elastic rather than viscous. At intermediate time scales mixed \vc\ response is observed. Therefore
the concept of a natural time of a material is important in characterizing the material as viscous
or elastic. The ratio between the material time scale and the time scale of the flow is indicated
by a non-dimensional number: the \Deborah\ or the \Weissenberg\ number \cite{BarnesbookHW1993}.

\vspace{0.2cm}

A common feature of \vc\ fluids is stress relaxation after a sudden shearing displacement where
stress overshoots to a maximum then starts decreasing exponentially and eventually settles to a
\steadys\ value. This phenomenon also takes place on cessation of steady shear flow where stress
decays over a finite measurable length of time. This reveals that \vc\ fluids are able to store and
release energy in contrast to inelastic fluids which react instantaneously to the imposed
deformation \cite{Deiberthesis1978, BirdbookAH1987, Larsonbook1988}. A defining characteristic of
\vc\ materials associated with stress relaxation is the relaxation time which may be defined as the
time required for the shear stress in a simple shear flow to return to zero under constant strain
condition. Hence for a \Hookean\ elastic solid the relaxation time is infinite, while for a \NEW\
fluid the relaxation of the stress is immediate and the relaxation time is zero. Relaxation times
which are infinite or zero are never realized in reality as they correspond to the mathematical
idealization of \Hookean\ elastic solids and \NEW\ liquids. In practice, stress relaxation after
the imposition of constant strain condition takes place over some finite non-zero time interval
\cite{OwensbookP2002}.

\vspace{0.2cm}

The complexity of \vy\ is phenomenal and the subject is notorious for being extremely difficult and
challenging. The constitutive equations for \vc\ fluids are much too complex to be treated in a
general manner. Further complications arise from the confusion created by the presence of other
phenomena such as wall effects and polymer-wall interactions, and these appear to be system
specific \cite{JonesH1979}. Therefore, it is doubtful that a general fluid model capable of
predicting all the flow responses of \vc\ systems with enough mathematical simplicity or
tractability can emerge in the foreseeable future \cite{Wissler1971, Deiberthesis1978,
ChhabraCM2001}. Understandably, despite the huge amount of literature composed in the last few
decades on this subject, the overwhelming majority of these studies have investigated very simple
cases in which substantial simplifications have been made using basic \vc\ models.

\vspace{0.2cm}

In the last few decades, a general consensus has emerged that in the flow of \vc\ fluids through
porous media elastic effects should arise, though their precise nature is unknown or controversial.
In porous media, \vc\ effects can be important in certain cases. When they are, the actual pressure
gradient will exceed the purely viscous gradient beyond a critical flow rate, as observed by
several investigators. The normal stresses of high molecular polymer solutions can explain in part
the high flow resistance encountered during \vc\ flow through porous media. It is believed that the
very high normal stress differences and \TR\ ratios associated with polymeric fluids will produce
increasing values of apparent viscosity when the flow channels in the porous medium are of rapidly
changing cross section.

\vspace{0.2cm}

Important aspects of \nNEW\ flow in general and \vc\ flow in particular through porous media are
still presenting serious challenge for modeling and quantification. There are intrinsic
difficulties in characterizing \nNEW\ effects in the flow of polymer solutions and the complexities
of the local geometry of the porous medium. This geometry gives rise to a complex and pore space
dependent flow field in which shear and extension coexist in various proportions that cannot be
quantified. Flows through porous media cannot be classified as pure shear flows as the \convdiv\
passages impose a predominantly extensional flow fields especially at high flow rates. The
extension viscosity of many \nNEW\ fluids also increases dramatically with the extension rate. As a
consequence, the relationship between the pressure drop and flow rate very often do not follow the
observed \NEW\ and inelastic \nNEW\ trend. Further complications arise from the fact that for
complex fluids the stress depends not only on whether the flow is a shearing, extensional, or mixed
type, but also on the whole history of the velocity gradient \cite{MarshallM1967, DaubenM1967,
Larsonbook1999, PearsonT2002, MendesN2002, PlogK}.

\subsection{Important Aspects for Flow in Porous Media} \label{ImportantAspects}
Strong experimental evidence indicates that the flow of \vc\ fluids through packed beds can exhibit
rapid increases in the pressure drop, or an increase in the apparent viscosity, above that expected
for a comparable purely viscous fluid. This increase has been attributed to the extensional nature
of the flow field in the pores caused by the successive expansions and contractions that a fluid
element experiences as it traverses the pore space. Although the flow field at pore level is not an
ideal extensional flow due to the presence of shear and rotation, the increase in flow resistance
is referred to as an \exThik\ effect \cite{DeiberS1981, ThienK1987, PilitsisB1989, PlogK}. In this
regard, two major interrelated aspects have strong impact on the flow through porous media. These
are extensional flow and \convdiv\ geometry.

\subsubsection{Extensional Flow}\label{ExtensionalFlow}
One complexity in the flow in general and through porous media in particular arises from the
coexistence of shear and extensional components, sometimes with the added complication of inertia.
Pure shear or elongational flow is the exception in practical situations. In most situations mixed
flow occurs where deformation rates have components parallel and perpendicular to the principal
flow direction. In such flows, the elongational components may be associated with the \convdiv\
flow paths \cite{Sorbiebook1991, BarnesbookHW1993}. A general consensus has emerged recently that
the flow through packed beds has a substantial extensional component and typical polymer solutions
exhibit strain hardening in extension, which is one of the main factors for the reported dramatic
increases in pressure drop. Thus in principle the shear viscosity alone is inadequate to explain
the observed excessive pressure gradients. It is interesting therefore to know the relative
importance of elastic and viscous effects or the relationship between normal and shear stresses for
different shear rates \cite{CarreaubookKC1997, ChhabraCM2001}.

\vspace{0.2cm}

An extensional or elongational flow is one in which fluid elements are subjected to extensions and
compressions without being rotated or sheared. The study of the extensional flow in general as a
relevant variable has only received attention in the last few decades with the realization that
extensional flow is of significant relevance in many practical situations. Before that, rheology
was largely dominated by shear flow. The historical convention of matching only shear flow data
with theoretical predictions in constitutive modeling should be rethought in those areas of
interest where there is a large extensional contribution. Extensional flow experiments can be
viewed as providing critical tests for any proposed constitutive equations \cite{Deiberthesis1978,
ChengH1984, BarnesbookHW1993, Larsonbook1999}.

\vspace{0.2cm}

Extensional flow is fundamentally different from shear flow. The material property characterizing
the extensional flow is not the viscosity but the extensional viscosity. The behavior of the
extensional viscosity function is often qualitatively different from that of the shear viscosity.
For example, highly elastic polymer solutions that possess a viscosity that decreases monotonically
in shear often exhibit an extensional viscosity that increases dramatically with extension rate.
Thus, while the shear viscosity is \shThin, the extensional viscosity is \exThik\
\cite{BarnesbookHW1993, Larsonbook1999}. The extensional or elongational viscosity $\exVis$, also
called \TR\ viscosity, is defined as the ratio of tensile stress to the extension rate under steady
flow condition where both these quantities attain constant values. Mathematically, it is given by
\begin{equation}\label{ExtensionViscosity}
    \exVis = \frac{\sS_{E}}{\elongR}
\end{equation}
where $\sS_{E}$ ($= \sS_{11} - \sS_{22}$) is the normal stress difference and $\elongR$ is the
extension rate. The stress $\sS_{11}$ is in the direction of the extension while $\sS_{22}$ is in a
direction perpendicular to the extension \cite{Lodgebook1964, BarnesbookHW1993}.

\vspace{0.2cm}

Polymeric fluids show a non-constant extensional viscosity in steady and unsteady extensional flow.
In general, extensional viscosity is a function of the extensional strain rate, just as the shear
viscosity is a function of shear rate. It is far more difficult to measure the extensional
viscosity than the shear viscosity. There is therefore a gulf between the strong desire to measure
extensional viscosity and the likely expectation of its fulfilment \cite{Larsonbook1988,
BarnesbookHW1993}. A major difficulty that hinders the progress in this field is that it is
difficult to achieve steady elongational flow and quantify it precisely. Despite the fact that many
techniques have been developed for measuring the elongational flow properties, these techniques
failed so far to produce consistent outcome as they generate results which can differ by several
orders of magnitude. This indicates that these results are dependent on the method and instrument
of measurement. This is highlighted by the view that the extensional viscometers provide
measurements of an extensional viscosity rather than the extensional viscosity. The situation is
made more complex by the fact that it is almost impossible to generate a pure extensional flow
since a shear component is always present in real flow situations, and this makes the measurements
doubtful and inconclusive \cite{BirdbookAH1987, BarnesbookHW1993, CarreaubookKC1997,
ChhabrabookR1999, Larsonbook1999, MendesN2002}.

\vspace{0.2cm}

For \NEW\ and inelastic \nNEW\ fluids the extensional viscosity is a constant that only depends on
the type of extensional deformation. Moreover, the viscosity measured in a shear flow can be used
to predict the viscosity in other types of deformation. For example, in a simple uniaxial
extensional flow of a \NEW\ fluid the following relationship is satisfied
\begin{equation}\label{exVisNewt}
    \exVis = 3 \lVis
\end{equation}

For \vc\ fluids the flow behavior is more complex and the extensional viscosity, like the shear
viscosity, depends on both the strain rate and the time following the onset of straining. The
rheological behavior of a complex fluid in extension is often very different from that in shear.
Polymers usually have extremely high extensional viscosities which can be orders of magnitude
higher than those expected on the basis of \NEW\ model. Moreover, the extensional viscosities of
elastic polymer solutions can be thousands of times greater than their shear viscosities. To
measure the departure of the ratio of extensional to shear viscosity from its \NEW\ equivalent, the
rheologists have introduced what is known as the \TR\ ratio which is usually defined as
\begin{equation}\label{Trouton}
    \Tr = \frac{\exVis(\elongR)}{\sVis(\sqrt{3}\elongR)}
\end{equation}

For elastic liquids, the \TR\ ratio is expected to be always greater than or equal to 3, with the
value 3 only attained at vanishingly small strain rates. These liquids are known for having very
high \TR\ ratios that can be as high as l0$^4$. This conduct is expected especially when the fluid
combines \shThin\ with \tenThik. However, even for the fluids that demonstrate \tenThin\ the
associated \TR\ ratios usually increase with strain rate and are still significantly in excess of
the inelastic value of three \cite{CrochetW1983, ChengH1984, Larsonbook1988, BarnesbookHW1993,
Larsonbook1999, OwensbookP2002}.


\begin{figure}[!t]
  \centering{}
  \includegraphics
  [scale=0.62]
  {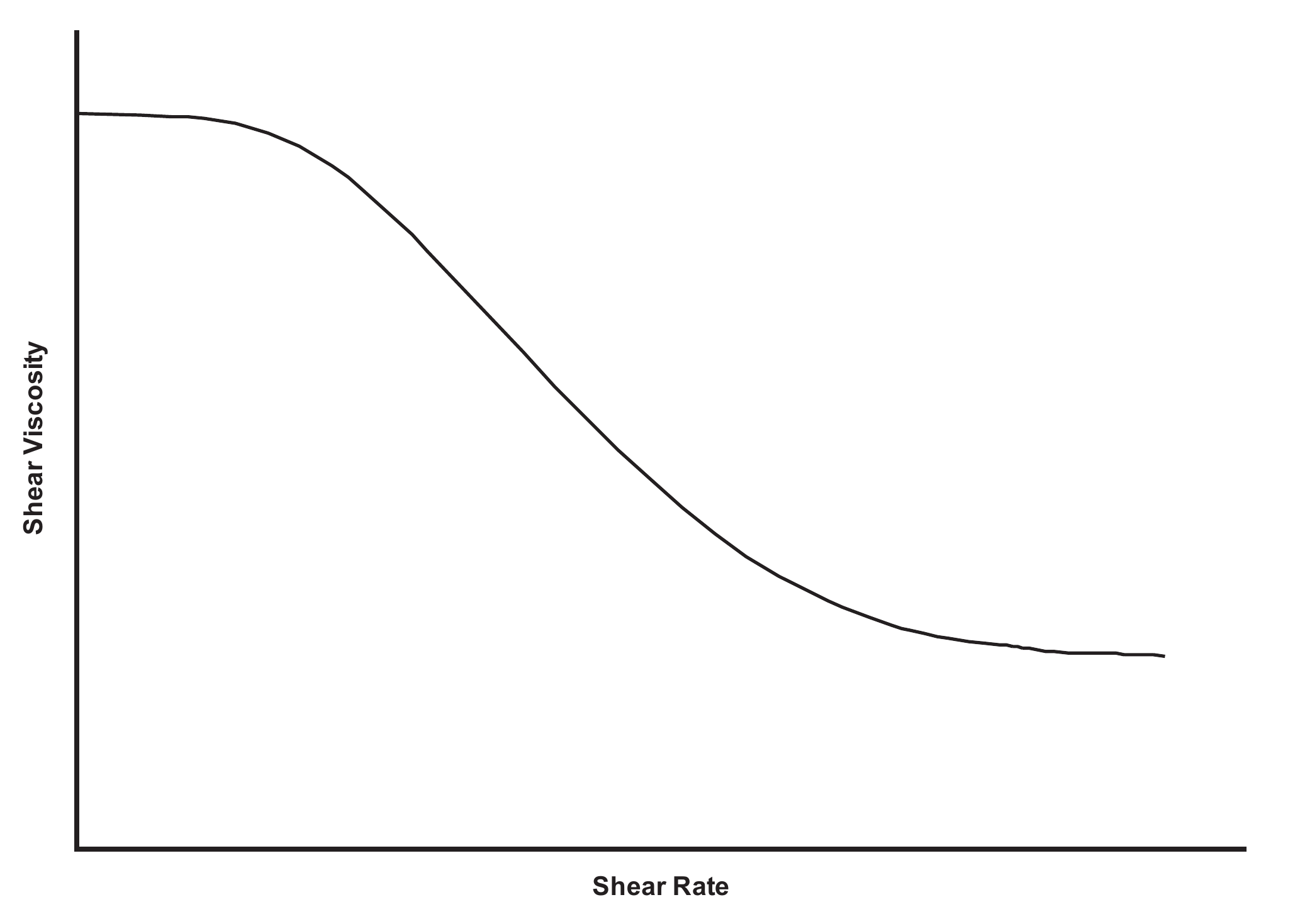}
  \caption[Typical behavior of shear viscosity $\sVis$ as a function of shear rate $\sR$ in shear flow on log-log scales]
  {Typical behavior of shear viscosity $\sVis$ as a function of shear rate $\sR$ in shear flow on log-log scales.}
  \label{ShearViscosity}
\end{figure}


\vspace{0.5cm}

\begin{figure}[!t]
  \centering{}
  \includegraphics
  [scale=0.62]
  {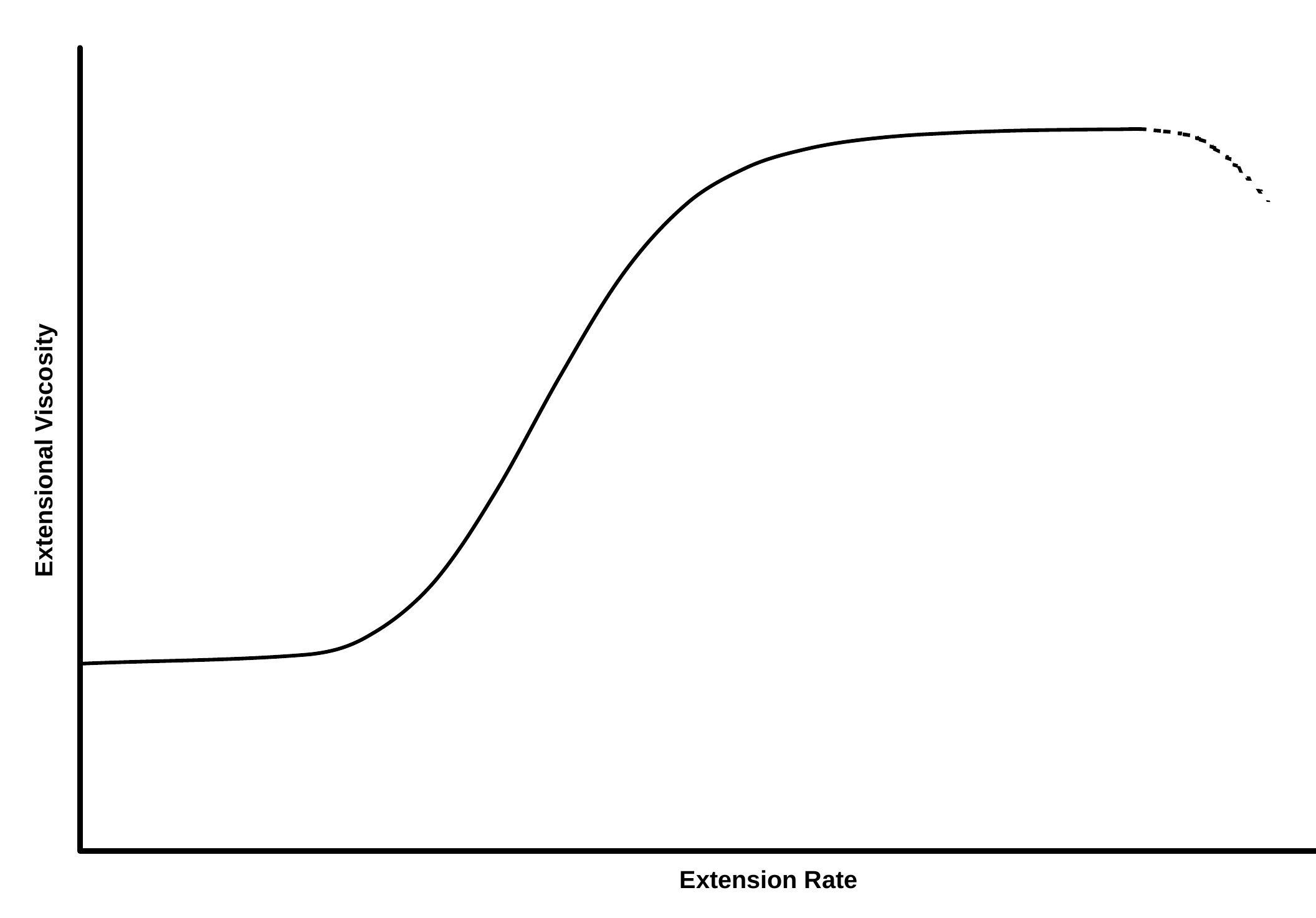}
  \caption[Typical behavior of extensional viscosity $\exVis$ as a function of extension rate $\elongR$ in extensional flow on log-log scales]
  {Typical behavior of extensional viscosity $\exVis$ as a function of extension rate $\elongR$ in extensional flow on log-log scales.}
  \label{ExtensionViscosity}
\end{figure}

\vspace{0.2cm}

Figures (\ref{ShearViscosity}) and (\ref{ExtensionViscosity}) compare the shear viscosity to the
extensional viscosity at isothermal condition for a typical \vc\ fluid at various shear and
extension rates. As seen in Figure (\ref{ShearViscosity}), the shear viscosity curve can be divided
into three regions. For low-shear rates, compared to the time scale of the fluid, the viscosity is
approximately constant. It then equals the so called zero shear rate viscosity. After this initial
plateau the viscosity rapidly decreases with increasing shear rate, and this behavior is described
as \shThin. However, there are some materials for which the reverse behavior is observed, that is
the shear viscosity increases with shear rate giving rise to \shThik. For high shear rates the
viscosity often approximates a constant value again. The constant viscosity extremes at low and
high shear rates are known as the lower and upper \NEW\ plateau, respectively \cite{BirdbookAH1987,
Larsonbook1988, BarnesbookHW1993}.

\vspace{0.2cm}

In Figure (\ref{ExtensionViscosity}) the typical behavior of the extensional viscosity, $\exVis$,
of a \vc\ fluid as a function of extension rate, $\elongR$, is depicted. As seen, the extensional
viscosity curve can be divided into several regions. At low extension rates the extensional
viscosity maintains a constant value known as the zero extension rate extensional viscosity. This
is usually three times the zero shear rate viscosity, just as for a \NEW\ fluid. For somewhat
larger extension rates the extensional viscosity increases with increasing extension rate. Some
\vc\ fluids behave differently, that is their extensional viscosity decreases on increasing
extension rate in this regime. A fluid for which $\exVis$ increases with increasing $\elongR$ is
said to be \tenThik, whilst if $\exVis$ decreases with increasing $\elongR$ it is described as
\tenThin. For even higher extension rates the extension viscosity reaches a maximum constant value.
If the extension rate is increased once more the extensional viscosity may decrease again. It
should be remarked that two polymeric liquids that may have essentially the same behavior in shear
can show a different response in extension \cite{BirdbookAH1987, BarnesbookHW1993, Larsonbook1999,
OwensbookP2002}.

\subsubsection{\ConvDiv\ Geometry} \label{ConvergingDiverging}
An important aspect that characterizes the flow in porous media and makes it distinct from bulk is
the presence of \convdiv\ flow paths. This geometric factor significantly affects the flow and
accentuate elastic responses. Several arguments have been presented in the literature to explain
the effect of \convdiv\ geometry on the flow behavior. One argument is that \vc\ flow in porous
media differs from that of \NEW\ flow, primarily because the \convdiv\ nature of flow in porous
media gives rise to normal stresses which are not solely proportional to the local shear rate. A
second argument is that any geometry that involves a diameter change will generate a flow field
with elongational characteristics, and hence the flow field in porous media involves both shear and
elongational components with elastic responses. A third argument suggests that elastic effects are
expected in the flow through porous media because of the acceleration and deceleration of the fluid
in the interstices of the bed upon entering and leaving individual pores \cite{TalwarK1992,
GogartyLF1972, ParkHB1973}.

\vspace{0.2cm}

Despite this diversity, there is a general consensus that in porous media the \convdiv\ nature of
the flow paths brings out both the extensional and shear properties of the fluid. The principal
mode of deformation to which a material element is subjected as the flow converges into a
constriction involves both a shearing of the material element and a stretching or elongation in the
direction of flow, while in the diverging portion the flow involves both shearing and compression.
The actual channel geometry determines the ratio of shearing to extensional contributions. In many
realistic situations involving \vc\ flows the extensional contribution is the most important of the
two modes. As porous media flow involves elongational flow components, the coil-stretch phenomenon
can also take place.

\vspace{0.2cm}

A consequence of the presence of \convdiv\ feature in porous media on flow modeling is that a
suitable model pore geometry is one having converging and diverging sections which can reproduce
the elongational nature of the flow. Despite the general success of the straight capillary tube
model with \NEW\ and inelastic \nNEW\ flows, its failure with elastic flow is remarkable. To
rectify the flaws of this model, the undulating tube and \convdiv\ channel were proposed in order
to include the elastic character of the flow. Various corrugated tube models with different simple
geometries have been used as a first step to model the effect of \convdiv\ geometry on the flow of
\vc\ fluids in porous media (e.g. \cite{MarshallM1967, DeiberS1981, SouvaliotisB1992}). Those
geometries include conically shaped sections, sinusoidal corrugation and abrupt expansions and
contractions. Some simplified forms of these geometries are displayed in Figure
(\ref{ConvDivGeom}). Similarly, a bundle of \convdiv\ tubes forms a better model for a porous
medium in \vc\ flow than the bundle of straight capillary tubes, as the presence of diameter
variations makes it possible to account for elongational contributions. Many investigators have
attempted to capture the role of successive \convdiv\ character of packed bed flow by numerically
solving the flow equations in conduits of periodically varying cross sections. Some of these are
\cite{Deiberthesis1978, PilitsisB1989, TalwarK1992, PodolsakTF1997, ChhabraCM2001}. Different
opinions on the success of these models can be found in the literature. With regards to modeling
\vc\ flow in regular or random networks of \convdiv\ capillaries, very little work has been done.

\subsection{\Vc\ Effects in Porous Media} \label{ViscoelasticEffects}
In packed bed flows, the main manifestation of \steadys\ \vc\ effects is the excess pressure drop
or \dilatancy\ behavior in different flow regimes above what is accounted for by shear viscosity of
the flowing liquid. Qualitatively, this behavior was attributed either to memory effects or to
extensional flow. However, both explanations have a place as long as the flow regime is considered.
Furthermore, the geometry of the porous media must be taken into account when considering elastic
responses \cite{GogartyLF1972, ChhabraCM2001}.

\vspace{0.2cm}

There is a general agreement that the flow of \vc\ fluids in packed beds results in a greater
pressure drop than what can be ascribed to the shear rate dependent viscosity. Fluids that exhibit
elasticity deviate from viscous flow above some critical velocity in porous flow. At low flow
rates, the pressure drop is determined largely by shear viscosity, and the viscosity and elasticity
are approximately the same as in bulk. As the flow rate gradually increases, \vc\ effects begin to
appear in various flow regimes. Consequently, the \insitu\ rheological characteristics become
significantly different from those in bulk as elasticity dramatically increases showing strong
\dilatant\ behavior. Although experimental evidence for \vc\ effects is convincing, an equally
convincing theoretical analysis is not available. One general argument is that when the fluid
suffers a significant deformation in a time comparable to the relaxation time of the fluid, elastic
effects become important \cite{Wissler1971, GogartyLF1972, CarreaubookKC1997, ChhabraCM2001}.

\vspace{0.2cm}

The complexity of \vc\ flow in porous media is aggravated by the possible occurrence of other
non-elastic phenomena which have similar effects as \vy. These phenomena include \adsorption\ of
the polymers on the capillary walls, mechanical \retention\ and partial pore blockage. All these
effects also lead to pressure drops well in excess to that expected from the shear viscosity
levels. Consequently, the interpretation of many observed effects is controversial. Some authors
may interpret some observations in terms of partial pore blockage whereas others insist on \nNEW\
effects including \vy\ as an explanation. However, none of these have proved to be completely
satisfactory. Moreover, no one can rule out the possibility of simultaneous occurrence of these
elastic and inelastic phenomena with further complication and confusion. An interesting fact is the
observation made by several researchers, e.g. Sadowski \cite{Sadowskithesis1963}, that reproducible
results can be obtained only in constant flow rate experiments because at constant pressure drop
the velocity kept decreasing. This kind of observation indicates deposition of polymer on the solid
surface by one mechanism or another, and cast doubt on some explanations which involve elasticity.
At constant flow rate the increased pressure drop seems to provide the necessary force to keep a
reproducible portion of the flow channels open \cite{DaubenM1967, Savins1969, Dullien1975,
ChhabraCM2001}.

\vspace{0.2cm}

There are three principal \vc\ effects that have to be accounted for in the investigation of \vy:
\trans\ time-dependence, \steadys\ time-dependence and \dilatancy\ at high flow rates.

\subsubsection{\Trans\ Time-Dependence} \label{TransientTimeDependence}
\Trans\ \timedep\ \vc\ behavior has been observed in bulk on the startup and cessation of processes
involving the displacement of \vc\ materials, and on a sudden change of rate or reversal in the
direction of deformation. During these \trans\ states, there are frequently overshoots and
undershoots in stress as a function of time which scale with strain. However, if the fluid is
strained at a constant rate, these transients die out in the course of time, and the stress
approaches a \steadys\ value that depends only on the strain rate. Under initial flow conditions
stresses can reach magnitudes which are substantially more important than their \steadys\ values,
whereas the relaxation on a sudden cessation of strain can vary substantially in various
circumstances \cite{BirdbookAH1987, CarreaubookKC1997, Larsonbook1999, BautistaSPM1999,
Tannerbook2000}.

\vspace{0.2cm}

\Trans\ responses are usually investigated in the context of bulk rheology, despite the fact that
there is no available theory that can predict this behavior quantitatively. As a consequence of
this restraint to bulk, the literature of \vc\ flow in porous media is almost entirely dedicated to
the \steadys\ situation with hardly any work on the \insitu\ \timedep\ \vc\ flows. However, \trans\
behavior is expected to occur in similar situations during the flow through porous media. One
reason for this gap is the absence of a proper theoretical structure and the experimental
difficulties associated with these flows \insitu. Another possible reason is that the \insitu\
\trans\ flows may have less important practical applications.

\subsubsection{\SteadyS\ Time-Dependence} \label{SteadyTimeDependence}
By this, we mean the effects that arise in the \steadys\ flow due to time dependency, as
time-dependence characteristics of the \vc\ material must have an impact on the \steadys\ conduct.
Indeed, elastic effects do occur in \steadys\ flow through porous media because of the
time-dependence nature of the flow at pore level. Depending on the time scale of the flow, \vc\
materials may show viscous or elastic behavior. The particular response in a given process depends
on the time scale of the process in relation to a natural time of the material. With regard to this
time scale dependency of \vc\ flow process at pore level, the fluid relaxation time and the rate of
elongation or contraction that occurs as the fluid flows through a channel or pore with varying
cross sectional area should be used to characterize and quantify \vc\ behavior \cite{MarshallM1967,
GogartyLF1972, HirasakiP1974, BarnesbookHW1993}.

\vspace{0.2cm}

Sadowski and Bird \cite{SadowskiB1965} analyzed the viscometric flow problem in a long straight
circular tube and argued that in such a flow no \timedep\ elastic effects are expected. However, in
a porous medium elastic effects may occur. As the fluid moves through a tortuous channel in the
porous medium, it encounters a capriciously changing cross-section. If the fluid relaxation time is
small with respect to the transit time through a contraction or expansion in a tortuous channel,
the fluid will readjust to the changing flow conditions and no elastic effects would be observed.
If, on the other hand, the fluid relaxation time is large with respect to the time to go through a
contraction or expansion, then the fluid will not accommodate and elastic effects will be observed
in the form of an extra pressure drop or an increase in the apparent viscosity. Thus the concept of
a ratio of characteristic times emerges as an ordering parameter in \vc\ flow through porous media.
This indicates the importance of the ratio of the natural time of a fluid to the duration time of a
process \cite{Savins1969}. It is noteworthy that this formulation relies on a twofold ordering
scheme and is valid for some systems and flow regimes. More elaborate formulation will be given in
conjunction with the intermediate plateau phenomenon.

\vspace{0.2cm}

One of the \steadys\ \vc\ phenomena observed in the flow through porous media and can be qualified
as a \timedep\ effect, among other possibilities such as \retention, is the intermediate plateau at
medium flow rates as demonstrated in Figure (\ref{VERheology2}). A possible explanation is that at
low flow rates before the appearance of the intermediate plateau the fluid behaves inelastically
like any typical \shThin\ fluid. This implies that in the low flow regime \vc\ effects are
negligible as the fluid can respond to the local state of deformation almost instantly, that is it
does not remember its past and hence behaves as a purely viscous fluid. As the flow rate increases,
a point will be reached where the solid-like characteristics of \vc\ materials begin to appear in
the form of an increased apparent viscosity as the time of process becomes comparable to the
natural time of fluid, and hence a plateau is observed. At higher flow rates the process time is
short compared to the natural time of the fluid and hence the fluid has no time to react as the
fluid is not an ideal elastic solid that reacts instantaneously. Since the process time is very
short, no overshoot will occur at the tube constriction as a measurable finite time is needed for
the overshoot to take place. The result is that no increase in the pressure drop will be observed
in this flow regime and the normal \shThin\ behavior is resumed with the eventual high flow rate
plateau \cite{MarshallM1967, LetelierSC2002}.

\vspace{0.2cm}

It is noteworthy that Dauben and Menzie \cite{DaubenM1967} have discussed a similar issue in
conjunction with the effect of \divconv\ geometry on the flow through porous media. They argued
that the process of expansion and contraction repeated many times may account in part for the high
pressure drops encountered during \vc\ flow in porous media. The fluid relaxation time combined
with the flow velocity determines the response of the \vc\ fluid in the suggested \divconv\ model.
Accordingly, it is possible that if the relaxation time is long enough or the fluid velocity high
enough the fluid can pass through the large opening before it has had time to respond to a stress
change imposed at the opening entrance, and hence the fluid would behave more as a viscous and less
like a \vc.

\subsubsection{\Dilatancy\ at High Flow Rates} \label{DilatancyatHighFlowRates}
The third distinctive feature of \vc\ flow is the \dilatant\ behavior in the form of excess
pressure losses at high flow rates as depicted in Figure (\ref{VERheology3}). Certain polymeric
solutions that exhibit \shThin\ in a viscometric flow seem to exhibit a \shThik\ response under
appropriate conditions during flow through porous media. At high flow rates, abnormal increases in
flow resistance that resemble a \shThik\ response have been observed in flow experiments involving
a variety of dilute to moderately concentrated solutions of high molecular weight polymers
\cite{Savins1969, Sorbiebook1991}. This phenomenon can be attributed to \stThik\ due to the
dominance of extensional over shear flow. At high flow rates, strong extensional flow effects do
occur and the extensional viscosity rises very steeply with increasing extension rate. As a result,
the non-shear terms become much larger than the shear terms. For \NEW\ fluids, the extensional
viscosity is just three times the shear viscosity. However, for \vc\ fluids the shear and
extensional viscosities often behave oppositely, that is while the shear viscosity is generally a
decreasing function of the shear rate, the extensional viscosity increases as the extension rate
increases. The consequence is that the pressure drop will be governed by \exThik\ and the apparent
viscosity rises sharply. Other possibilities such as physical \retention\ are less likely to take
place at these high flow rates \cite{DurstHI1987, MendesN2002}.

\vspace{0.4cm}

Finally, a glimpse on the literature of \vc\ flow in porous media will help to clarify the
situation and highlight some important contributions in this field. Sadowski and Bird
\cite{SadowskiB1965, Sadowski1965, Sadowskithesis1963} were the first to include elastic effects in
their rheological model to account for a departure of the experimental data in porous media from
the modified \Darcy's law \cite{Savins1969, Sorbiebook1991}. Wissler \cite{Wissler1971} was the
first to account quantitatively for the elongational stresses developed in \vc\ flow through porous
media \cite{TalwarK1992}. In this context, he presented a third order perturbation analysis of a
\vc\ fluid flow through a \convdiv\ channel. Gogarty \etal\ \cite{GogartyLF1972} developed a
modified form of the \nNEW\ \Darcy\ equation to predict the viscous and elastic pressure drop
versus flow rate relationships for the flow of elastic \carboxymethylcellulose\ (CMC) solutions in
beds of different geometries. Park \etal\ \cite{ParkHB1973} studied the flow of various polymeric
solutions through packed beds using several rheological models. In the case of one type of solution
at high \Reynolds\ numbers they introduced an empirical corrections based upon a pseudo \vc\
\Deborah\ number to improve the data fit.

\vspace{0.2cm}

In their investigation to the flow of polymers through porous media, Hirasaki and Pope
\cite{HirasakiP1974} modeled the \dilatant\ behavior by \vc\ properties of the polymer solution.
The additional \vc\ resistance to the flow, which is a function of \Deborah\ number, was modeled as
a simple elongational flow to represent the elongation and contraction that occur as the fluid
flows through a pore with varying cross sectional area. Deiber and Schowalter \cite{DeiberS1981,
Deiberthesis1978} used a circular tube with a radius which varies sinusoidally in the axial
direction as a first step toward modeling the flow channels of a porous medium. They concluded that
such a tube exhibits similar phenomenological aspects to those found for the flow of \vc\ fluids
through packed beds. Durst \etal\ \cite{DurstHI1987} highlighted the fact that the pressure drop of
a porous medium flow is only due to a small extent to the shear force term usually employed to
derive the \Kozeny-\Darcy\ law. For a more correct derivation, additional terms have to be taken
into account since the fluid is also exposed to elongational forces when it passes through the
porous media matrix. Chmielewski and coworkers \cite{ChmielewskiPJ1990, ChmielewskiJ1992,
ChmielewskiJ1993} investigated the elastic behavior of \polyisobutylene\ solutions flowing through
arrays of cylinders with various geometries. They recognized that the \convdiv\ geometry of the
pores in porous media causes an extensional flow component that may be associated with the
increased flow resistance for \vc\ liquids. Pilitsis \etal\ \cite{PilitsisSB1991} simulated the
flow of a \shThin\ \vc\ fluid through a periodically constricted tube using a variety of
constitutive rheological models, and the results were compared against the experimental data of
Deiber and Schowalter \cite{DeiberS1981}. It was found that the presence of the elasticity in the
mathematical model caused an increase in the flow resistance over the value calculated for the
viscous fluid. Talwar and Khomami \cite{TalwarK1992} developed a higher order Galerkin finite
element scheme for simulating two-dimensional \vc\ flow and successfully tested the algorithm with
the flow of \UpCoMa\ (UCM) and \OldroydB\ fluids in undulating tube. It was subsequently used to
solve the problem of transverse steady flow of UCM and \OldroydB\ fluids past an infinite square
arrangement of infinitely long, rigid, motionless cylinders.

\vspace{0.2cm}

Garrouch \cite{Garrouch1999} developed a generalized \vc\ model for polymer flow in porous media
analogous to \Darcy's law, and hence concluded a nonlinear relationship between the fluid velocity
and pressure gradient. The model accounts for polymer elasticity by using the longest relaxation
time, and accounts for polymer viscous properties by using an average porous medium \powlaw\
behavior index. Koshiba \etal\ \cite{KoshibaMNS2000} investigated the \vc\ flow through an
undulating channel as a model for flow paths in porous media and concluded that the stress in the
flow through the undulating channel should rapidly increase with increasing flow rate because of
the \stThik\ elongational viscosity. Khuzhayorov \etal\ \cite{KhuzhayorovAR2000} applied a
homogenization technique to derive a macroscopic filtration law for describing \trans\ linear \vc\
fluid flow in porous media. The results obtained in the particular case of the flow of an \Oldroyd\
fluid in a bundle of capillary tubes show that \vc\ behavior strongly differs from \NEW\ behavior.
Huifen \etal\ \cite{HuifenXDQX2001} developed a model for the variation of rheological parameters
along the seepage flow direction and constructed a constitutive equation for \vc\ fluids in which
the variation of the rheological parameters of polymer solutions in porous media is taken into
account. Mendes and Naccache \cite{MendesN2002} employed a simple theoretical approach to obtain a
constitutive relation for the flows of \exThik\ fluids through porous media. The pore morphology is
assumed to be composed of a bundle of periodically \convdiv\ tubes. Dolej$\check{\rm{s}}$ \etal\
\cite{DolejsCSD2002} presented a method for the pressure drop calculation for the flow of \vc\
fluids through fixed beds of particles. The method is based on the application of the modified
\Rabinowitsch-\Mooney\ equation together with the corresponding relations for consistency
variables.

\newpage

\section{\Thixotropy\ and \Rheopexy} \label{Thixotropy}
\Timedep\ fluids are defined to be those fluids whose viscosity depends on the duration of flow
under isothermal conditions. For these fluids, a \timeind\ \steadys\ viscosity is eventually
reached in steady flow situation. The time effect is often reversible though it may be partial,
that is the trend of the viscosity change is overturned in time when the stress is reduced. The
phenomenon is generally attributed to \timedep\ \thixotropic\ breakdown or \rheopectic\ buildup of
some particulate structure under relatively high stress followed by structural change in the
opposite direction for lower stress, though the exact mechanism may not be certain
\cite{Collyer1974, EscudierP1996, CarreaubookKC1997, Barnes1997, DullaertM2005}.

\vspace{0.2cm}

\Timedep\ fluids are generally divided into two main categories: \thixotropic\ (work softening)
whose viscosity gradually decreases with time at a constant shear rate, and \rheopectic\ (work
hardening or anti-\thixotropic\ or negative \thixotropic) whose viscosity increases under similar
circumstances without an overshoot which is a characteristic feature of \vy. However, it has been
proposed that \rheopexy\ and negative \thixotropy\ are different, and hence three categories of
\timedep\ fluids do exist \cite{ChengE1965, CarreaubookKC1997, OwensbookP2002}. It is noteworthy
that in this article we may rely on the context and use `\thixotropy' conveniently to indicate
non-elastic time-dependence in general where the meaning is obvious.

\vspace{0.2cm}

\Thixotropic\ fluids may be described as \shThin\ while the \rheopectic\ as \shThik, in the sense
that these effects take place on shearing, though they are effects of time-dependence. However, it
is proposed that \thixotropy\ invariably occurs in circumstances where the liquid is \shThin\ (in
the usual sense of viscosity decrease with increasing shear rate) while \rheopexy\ may be
associated with \shThik. This can be behind the occasional confusion between \thixotropy\ and
\shThin\ \cite{Mewis1979, BarnesbookHW1993, Barnes1997, BalmforthbookC2001}.

\vspace{0.2cm}

A substantial number of complex fluids display time-dependence phenomena, with \thixotropy\ being
more commonplace and better understood than \rheopexy. Various mathematical models have been
proposed to describe time-dependence behavior. These include microstructural, continuum mechanics,
and structural kinetics models. \Thixotropic\ and \rheopectic\ behaviors may be detected by the
hysteresis loop test, as well as by the more direct steady shear test. In the loop test the
substance is sheared cyclically and a graph of stress versus shear rate is obtained. A \timedep\
fluid should display a hysteresis loop the area of which is a measure of the degree of \thixotropy\
or \rheopexy\ and this may be used to quantify \timedep\ behavior \cite{ChengE1965, Collyer1974,
Mewis1979, ChhabrabookR1999, DullaertM2005}.

\vspace{0.2cm}

In theory, time-dependence effects can arise from \thixotropic\ structural change or from \vy. The
existence of these two different types of \timedep\ rheologies is generally recognized. Although it
is convenient to distinguish between these as separate types of phenomena, real fluids can exhibit
both types of rheology at the same time. Several physical distinctions between \vc\ and
\thixotropic\ time-dependence have been made. The important one is that while the time-dependence
of \vc\ fluids arises because the response of stresses and strains in the fluid to changes in
imposed strains and stresses respectively is not instantaneous, in \thixotropic\ fluids such
response is instantaneous and the \timedep\ behavior arises purely because of changes in the
structure of the fluid as a consequence of strain. While the mathematical theory of \vy\ has been
developed to an advanced level, especially on the continuum mechanical front, relatively little
work has been done on \thixotropy\ and \rheopexy. One reason is the lack of a comprehensive
framework to describe the dynamics of \thixotropy. This may partly explain why \thixotropy\ is
rarely incorporated in the constitutive equation when modeling the flow of \nNEW\ fluids. The
underlying assumption is that in these situations the \thixotropic\ effects have a negligible
impact on the resulting flow field, and this allows great mathematical simplifications
\cite{ChengE1965, JoyeP1971, BillinghamF1993, EscudierP1996, PearsonT2002, PritchardP2006}.


\begin{figure} [!t]
  \centering{}
  \includegraphics
  [scale=0.6]
  {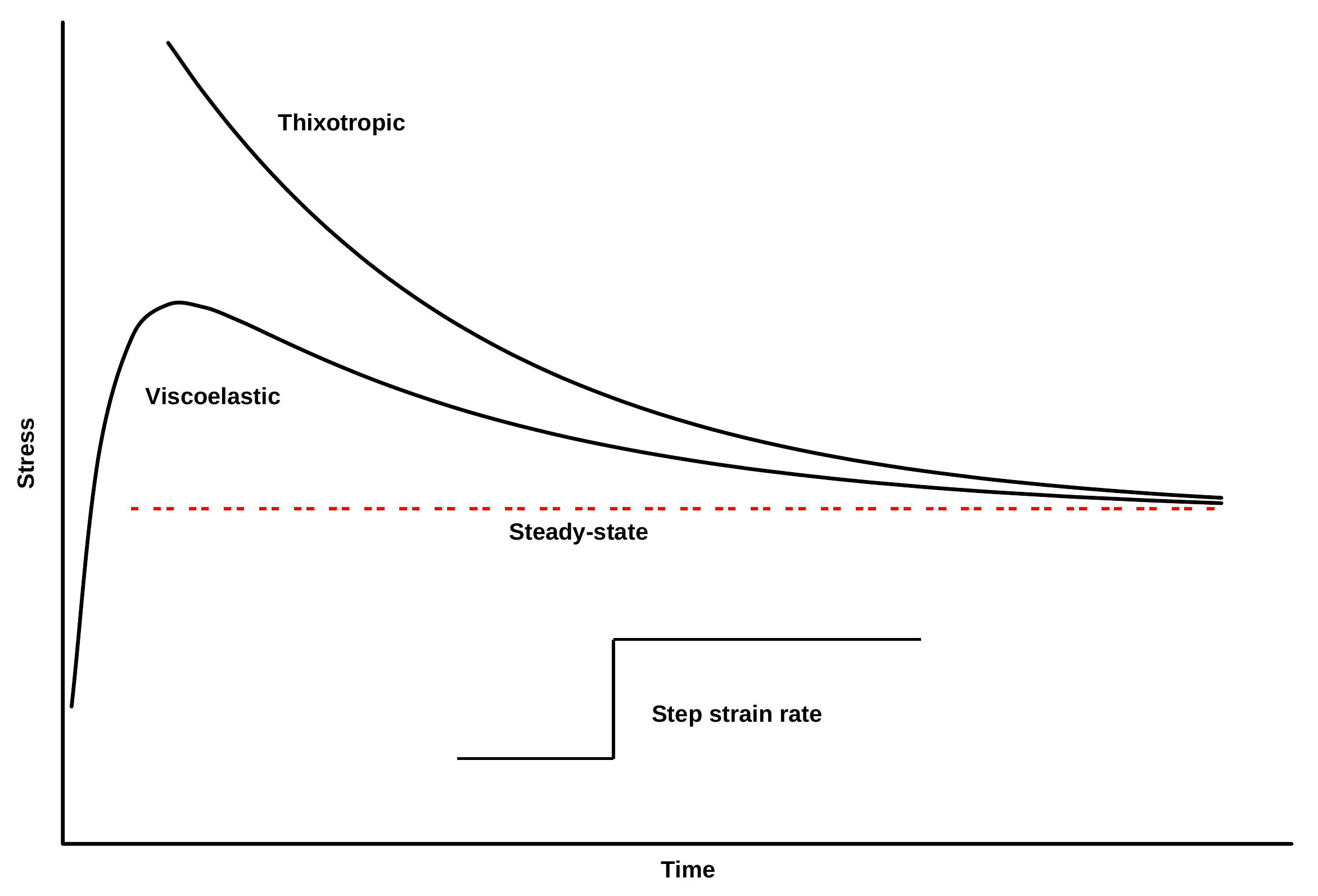}
  \caption[Comparison between time dependency in \thixotropic\ and \vc\ fluids following a step increase in strain rate]
  {Comparison between time dependency in \thixotropic\ and \vc\ fluids following a step increase in strain rate.}
  \label{ThixoVE}
\end{figure}


\vspace{0.2cm}

Several behavioral distinctions can be made to differentiate between \vy\ and \thixotropy. These
include the presence or absence of some characteristic elastic features such as recoil and normal
stresses. However, these signs may be of limited use in some practical situations involving complex
fluids where these two phenomena coexist. In Figure (\ref{ThixoVE}) the behavior of these two types
of fluids in response to step change in strain rate is compared. Although both fluids show signs of
dependency on the past history, the graph suggests that inelastic \thixotropic\ fluids do not
possess a memory in the same sense as \vc\ materials. The behavioral differences, such as the
absence of elastic effects or the difference in the characteristic time scale associated with these
phenomena, confirm this suggestion \cite{JoyeP1971, Mewis1979, Barnes1997, CarreaubookKC1997}.

\vspace{0.2cm}

\Thixotropy, like \vy, is a phenomenon that can appear in a large number of systems. The time scale
for \thixotropic\ changes is measurable in various materials including important commercial and
biological products. However, the investigation of \thixotropy\ has been hampered by several
difficulties. Consequently, the suggested \thixotropic\ models seem unable to present successful
quantitative description of \thixotropic\ behavior especially for the \trans\ state. In fact, even
the most characteristic property of \thixotropic\ fluids, i.e. the decay of viscosity or stress
under steady shear conditions, presents difficulties in modeling and characterization. The lack of
a comprehensive theoretical framework to describe \thixotropy\ is matched by a severe shortage on
the experimental side. One reason is the difficulties confronted in measuring \thixotropic\ systems
with sufficient accuracy. As a result, very few systematic data sets can be found in the literature
and this deficit hinders the progress in this field. Moreover, the characterization of the data in
the absence of an agreed-upon mathematical structure may be questionable \cite{JoyeP1971,
Godfrey1973, Mewis1979, EscudierP1996, DullaertM2005}.

\subsection{Modeling \TimeDep\ Flow in Porous Media} \label{SimulatingTDFPM}
In the absence of serious attempts to model \thixotropic\ rheology in porous media using the four
major modeling approaches (i.e. continuum, capillary bundle, numerical methods and network
modeling), very little can be said about this issue. It seems, however, that network modeling is
currently the best candidate for dealing with this task. In this context, There are three major
cases of simulating the flow of \timedep\ fluids in porous media

\begin{itemize}

\item The flow of strongly strain-dependent fluid in a porous medium that
is not very homogeneous. This case is very difficult to model because of the difficulty to track
the fluid elements in the pores and throats and determine their deformation history. Moreover, the
viscosity function is not well defined due to the mixing of fluid elements with various deformation
history in the individual pores and throats.

\item The flow of strain-independent or weakly strain-dependent fluid
through porous media in general. A possible strategy is to apply single \timedep\ viscosity
function to all pores and throats at each instant of time and hence simulating time development as
a sequence of \NEW\ states.

\item The flow of strongly strain-dependent fluid in a very homogeneous
porous medium such that the fluid is subject to the same deformation in all network elements. The
strategy for modeling this flow is to define an effective pore strain rate. Then using a very small
time step the strain rate in the next instant of time can be found assuming constant strain rate.
As the change in the strain rate is now known, a correction to the viscosity due to
strain-dependency can be introduced.

There are two possible problems in this strategy. The first is edge effects in the case of
injection from a reservoir because the deformation history of the fluid elements at the inlet is
different from the deformation history of the fluid elements inside the network. The second is that
considerable computing resources may be required if very small time steps are used.

\end{itemize}

\vspace{0.4cm}

The flow of \timedep\ fluids in porous media has not been vigorously investigated. Consequently,
very few studies can be found in the literature on this subject. One reason is that \timedep\
effects are usually investigated in the context of \vy. Another reason is the lack of a
comprehensive framework to describe the dynamics of \timedep\ fluids in porous media
\cite{ChengE1965, PearsonT2002, PritchardP2006}. Among the few examples that we found on this
subject is the investigation of Pritchard and Pearson \cite{PritchardP2006} of viscous fingering
instability during the injection of a \thixotropic\ fluid into a porous medium or a narrow
fracture. Another example is the study by Wang \etal\ \cite{WangHC2006} who examined \thixotropic\
effects in the context of heavy oil flow through porous media. Finally, some \thixotropic\ aspects
were modeled by Sochi \cite{Sochithesis2007} using network modeling approach as part of
implementing the \BauMan\ rheological model which incorporates \thixotropic\ as well as \vc\
effects.

\newpage

\section [Experimental Work] {Experimental Work on \NNEW\ Flow in
Porous Media} \label{Experimental}
For completion, we include a brief non-comprehensive list of some experimental work in the field of
\nNEW\ flow through porous media. This collection should provide a feeling of the nature and topics
of experimental research in the last few decades. The list is arranged in a chronological order

\begin{itemize}

\item Sadowski \cite{Sadowskithesis1963, Sadowski1965} conducted extensive
experimental work on the flow of polymeric solutions through packed beds. He used ten aqueous
polymeric solutions made from \Carbowax\ 20-M, \Elvanol\ 72-51, \Natrosol\ 250G and \Natrosol\ 250H
with various concentrations. The porous media were packed beds of lead shot or glass beads with
different properties. He used \Ellis\ as a rheological model for the fluids and introduced a
characteristic time to designate regions of behavior where elastic effects are important.

\item Marshall and Metzner \cite{MarshallM1967} used three polymeric
\nNEW\ solutions (\Carbopol\ 934, \polyisobutylene\ L100 and ET-597) to investigate the flow of
\vc\ fluids through porous media and determine a \Deborah\ number at which \vc\ effects first
become measurable. The porous medium which they used was a sintered bronze porous disk.

\item White \cite{White1968} conducted simple experiments on the flow of
a \hydroxyethylcellulose\ solution through a stratified bed using a \powlaw\ form of \Darcy's Law.

\item Gogarty \etal\ \cite{GogartyLF1972} carried out experiments on the
flow of polymer solutions of \carboxymethylcellulose\ (CMC) and \polyacrylamide\ through three
packed beds of glass beads with various size and a packed bed of sand.

\item Park \etal\ \cite{ParkHB1973, Parkthesis1972} experimentally studied
the flow of polymeric aqueous solutions of \polymethylcellulose\ (PMC) with different molecular
weights and concentrations through two packed beds of spherical uniform-in-size glass beads, coarse
and fine. Several rheological models, including \Ellis\ and \HB, were used to characterize the
fluids.

\item Teeuw and Hesselink \cite{TeeuwH1980} carried out core
flow experiments to investigate the behavior of aqueous biopolymer solutions in porous media. They
used cylindrical plugs of \Bentheim, a clean outcrop sandstone of very uniform grain size and
structure, as porous media characterizing the solutions with \powlaw.

\item Vogel and Pusch \cite{VogelP1981} conducted flow experiments
on \sandp\ using three polymeric solutions: a \polysaccharide, a \hydroxyethylcellulose\ and a
\polyacrylamide.

\item Al-Fariss and Pinder \cite{AlfarissP1984} conducted experimental
work on waxy oils modeled as \HB\ fluids. The porous media consist of two packed beds of sand
having different dimensions, porosity, permeability and grain size.

\item Durst \etal\ \cite{DurstHI1987} verified aspects of their
theoretical \nNEW\ investigation by experimental work on the flow of dilute polymer solutions
through porous media of packed spheres.

\item Sorbie \etal\ \cite{SorbiePC1987} conducted an extensive
experimental and theoretical investigation on the single-phase flow of \Xanthan/tracer slugs in a
consolidated sandstone using \Carreau\ rheology. The phenomena studied include polymer/tracer
dispersion, excluded volume effects, polymer adsorption, and viscous fingering.

\item Cannella \etal\ \cite{CannellaHS1988} experimentally investigated
the flow of \Xanthan\ solutions through \Berea\ sandstone and carbonate cores in the presence and
absence of residual oil. They used modified \powlaw\ and \Carreau\ rheologies in their theoretical
analysis.

\item Chmielewski \etal\ \cite{ChmielewskiPJ1990, ChmielewskiJ1992,
ChmielewskiJ1993} conducted experiments to investigate the elastic behavior of the flow of
\polyisobutylene\ solutions through arrays of cylinders with various geometries.

\item Chhabra and Srinivas \cite{ChhabraS1991} conducted
experimental work investigating the flow of \powlaw\ liquids through beds of non-spherical
particles using the \Ergun\ equation to correlate the resulting values of \fricfact. They used beds
of three different types of packing materials (two sizes of \Raschig\ rings and one size of gravel
chips) as porous media and solutions of \carboxymethylcellulose\ as \shThin\ liquids.

\item Fletcher \etal\ \cite{FletcherFLBSG1991} investigated the
flow of biopolymer solutions of \Flocon\ 4800MXC, \Xanthan\ CH-100-23, and \scleroglucan\ through
\Berea\ and \Clashach\ cores utilizing \Carreau\ rheological model to characterize the fluids.

\item Hejri \etal\ \cite{HejriWG1991} investigated the rheological
behavior of biopolymer solutions of \Flocon\ 4800MX characterized by \powlaw\ in \sandp\ cores.

\item Sabiri and Comiti \cite{SabiriC1995} investigated the flow of \nNEW\
purely viscous fluids through packed beds of spherical particles, long cylinders and very flat
plates. They used aqueous polymeric solutions of \carboximethylcellulose\ sodium salt. The
rheological behavior of these solutions was represented by a series of \powlaw\ type equations.

\item Saunders \etal\ \cite{SaundersLB1999} carried experimental studies
on \vc\ compression of resin-impregnated fibre cloths. The experiments included a type of
plain-weave cloth and two types of resin, an epoxy behaving approximately as a \NEW\ fluid and a
polyester following \powlaw\ \nNEW\ behavior.

\item Chase and Dachavijit \cite{ChaseD2003} performed experimental
research on aqueous solutions of \Carbopol\ 941 with various concentrations modeled as \Bingham\
fluids. The porous medium was a packed column of spherical glass beads having a narrow size
distribution.

\item Kuzhir \etal\ \cite{KuzhirBBV2003} investigated the flow of a
\Bingham\ magneto-rheological (MR) fluid through different types of porous medium which include
bundles of cylinders and beds of magnetic and non-magnetic spheres.

\item D'Angelo \etal\ \cite{DangeloFCR2003} used radioactive techniques
to study the dispersion and \retention\ phenomena of \nNEW\ \scleroglucan\ polymeric solutions in a
porous medium consisting of an acrylic cylinder filled with compacted glass microspheres.

\item Balhoff and Thompson \cite{BalhoffT2006, Balhoffthesis2005} carried
out a limited amount of experimental work (single dataset) on the flow of \guar\ gum solution in a
packed bed of glass beads. \Ellis\ rheology was used to characterize the fluid.

\item Perrin \etal\ \cite{PerrinTSC2006} conducted experimental work on
the flow of hydrolyzed \polyacrylamide\ (HPAM) solutions in two-dimensional etched silicon wafer
micromodels as idealizations of porous media. The results from these experiments were analyzed by
pore-scale network modeling incorporating the \nNEW\ fluid mechanics of a \Carreau\ fluid.

\item Wang \etal\ \cite{WangHC2006} carried out experimental work on dead
oils obtained from four wells in the Chinese Zaoyuan oilfield by injecting oil into \sandp\ cores.
The investigation included \yields, \vc\ and \thixotropic\ aspects using \HB\ model.

\item Schevena \etal\ \cite{SchevenSAHJG2007} conducted experimental
research on \NEW\ and \nNEW\ flows through rocks and bead packs. The experiments involved the use
of \shThin\ \Xanthan\ solutions and packed beds of near-monodisperse \Ballotini\ beads.

\end{itemize}

\newpage

\section [Conclusions] {Conclusions} \label{Conclusions}
In the context of fluid flow, `\nNEW' is a generic term that incorporates a variety of phenomena.
These phenomena are highly complex and require sophisticated mathematical modeling techniques for
proper description. Further complications are added when considering the flow through porous media.
So far no general methodology that can deal with all cases of \nNEW\ flow has been developed.
Moreover, only modest success has been achieved by any one of these methodologies. This situation
is not expected to change substantially in the foreseeable future, and many challenges are still
waiting to overcome. In the absence of a general approach that is suitable for all situations, a
combination of all available approaches is required to tackle \nNEW\ challenges.

\vspace{0.2cm}

Currently, network modeling is the most realistic choice for modeling \nNEW\ flow in porous media.
While this approach is capable of predicting the general trend in many situations, it is still
unable to account for all complexities and make precise predictions in all cases. The way forward
is to improve the modeling strategies and techniques. One requirement is the improvement of pore
space definition. While modeling the \convdiv\ feature of the pore space with idealized geometry is
a step forward, it is not enough. This feature is only one factor in the making of the complex
structure of flow channels. The actual geometry and topology in porous media is much more complex
and should be fully considered for successful modeling of flow field. Including more physics in the
flow description at pore level is another requirement for reaching the final objective.

\newpage

\section[Terminology of Flow and \Vy]{Appendix A} \label{AppVE}

{ 
\renewcommand{\thefootnote}{\fnsymbol{footnote}}

{\Large \bf Terminology of Flow and \Vy\footnote{In preparing this Appendix, we consulted most of
our references. The main ones are \cite{Skellandbook1967, Boger1977, BirdbookAH1987,
Larsonbook1988, Whitebook1991, BarnesbookHW1993, Larsonbook1999, Tannerbook2000, OsP2004}.} }

\vspace{0.5cm}

\noindent A tensor is an array of numbers which transform according to certain rules under
coordinate change. In a three-dimensional space, a tensor of order $n$ has $3^{n}$ components which
may be specified with reference to a given coordinate system. Accordingly, a scalar is a zero-order
tensor and a vector is a first-order tensor.

\vspace{0.2cm}

A stress is defined as a force per unit area. Because both force and area are vectors, considering
the orientation of the normal to the area, stress has two directions associated with it instead of
one as with vectors. This indicates that stress should be represented by a second-order tensor,
given in Cartesian coordinate system by
\begin{equation}\label{strTens}
    \sTen = \left(
               \begin{array}{ccc}
                 \sTenC_{xx} & \sTenC_{xy} & \sTenC_{xz} \\
                 \sTenC_{yx} & \sTenC_{yy} & \sTenC_{yz} \\
                 \sTenC_{zx} & \sTenC_{zy} & \sTenC_{zz} \\
               \end{array}
             \right)
\end{equation}
where $\sTenC_{ij}$ is the stress in the $j$-direction on a surface normal to the $i$-axis. A shear
stress is a force that a flowing liquid exerts on a surface, per unit area of that surface, in the
direction parallel to the flow. A normal stress is a force per unit area acting normal or
perpendicular to a surface. The components with identical subscripts represent normal stresses
while the others represent shear stresses. Thus $\sTenC_{xx}$ is a normal stress acting in the
$x$-direction on a face normal to $x$-direction, while $\sTenC_{yx}$ is a shear stress acting in
the $x$-direction on a surface normal to the $y$-direction, positive when material at greater $y$
exerts a shear in the positive $x$-direction on material at lesser $y$. Normal stresses are
conventionally positive when tensile. The stress tensor is symmetric that is $\sTenC_{ij} =
\sTenC_{ji}$ where $i$ and $j$ represent $x$, $y$ or $z$. This symmetry is required by angular
momentum considerations to satisfy equilibrium of moments about the three axes of any fluid
element. This means that the state of stress at a point is determined by six, rather than nine,
independent stress components.

\vspace{0.2cm}

\Vc\ fluids show normal stress differences in steady shear flows. The first normal stress
difference $\fNSD$ is defined as
\begin{equation}\label{fNSD}
    \fNSD = \sTenC_{11} - \sTenC_{22}
\end{equation}
where $\sS_{11}$ is the normal stress component acting in the direction of flow and $\sS_{22}$ is
the normal stress in the gradient direction. The second normal stress difference $\sNSD$ is defined
as
\begin{equation}\label{sNSD}
    \sNSD = \sTenC_{22} - \sTenC_{33}
\end{equation}
where $\sS_{33}$ is the normal stress component in the indifferent direction. The magnitude of
$\sNSD$ is in general much smaller than $\fNSD$. For some \vc\ fluids, $\sNSD$ may be virtually
zero. Often not the first normal stress difference $\fNSD$ is given, but a related quantity: the
first normal stress coefficient. This coefficient is defined by $\Psi_{1} = \frac{\fNSD}{\sR^{2}}$
and decreases with increasing shear rate. Different conventions about the sign of the normal stress
differences do exist. However, in general $\fNSD$ and $\sNSD$ show opposite signs. \Vc\ solutions
with almost the same viscosities may have very different values of normal stress differences. The
presence of normal stress differences is a strong indication of \vy, though some associate these
properties with non-\vc\ fluids.

\vspace{0.2cm}

The total stress tensor, also called Cauchy stress tensor, is usually divided into two parts:
hydrostatic stress tensor and extra stress tensor. The former represents the hydrostatic pressure
while the latter represents the shear and extensional stresses caused by the flow. In equilibrium
the pressure reduces to the hydrostatic pressure and the extra stress tensor $\sTen$ vanishes. The
extra stress tensor is determined by the deformation history and has to be specified by the
constitutive equation of the particular fluid.

\vspace{0.2cm}

The rate of strain or rate of deformation tensor is a symmetric second order tensor which gives the
components, both extensional and shear, of the strain rate. In Cartesian coordinate system it is
given by:
\begin{equation}\label{rateStrTens}
    \rsTen = \left(
               \begin{array}{ccc}
                 \rsTenC_{xx} & \rsTenC_{xy} & \rsTenC_{xz} \\
                 \rsTenC_{yx} & \rsTenC_{yy} & \rsTenC_{yz} \\
                 \rsTenC_{zx} & \rsTenC_{zy} & \rsTenC_{zz} \\
               \end{array}
             \right)
\end{equation}
where $\rsTenC_{xx}$, $\rsTenC_{yy}$ and $\rsTenC_{zz}$ are the extensional components while the
others are the shear components. These components are given by:
\begin{eqnarray} \label{rsTenComp}
  \hspace{-2.0cm}
  \rsTenC_{xx} = 2 \frac{\partial \vC_{x}}{\partial x}  \verb|       |
  \rsTenC_{yy} = 2 \frac{\partial \vC_{y}}{\partial y}  \verb|       |
  \rsTenC_{zz} = 2 \frac{\partial \vC_{z}}{\partial z}  \verb|       |
  \hspace{-4.0cm}
  \nonumber \\
  \rsTenC_{xy} = \rsTenC_{yx} =
  \frac{\partial \vC_{x}}{\partial y} + \frac{\partial \vC_{y}}{\partial x}
  \nonumber \\
  \rsTenC_{yz} = \rsTenC_{zy} =
  \frac{\partial \vC_{y}}{\partial z} + \frac{\partial \vC_{z}}{\partial y}
  \nonumber \\
  \rsTenC_{xz} = \rsTenC_{zx} =
  \frac{\partial \vC_{x}}{\partial z} + \frac{\partial \vC_{z}}{\partial x}
\end{eqnarray}
where $\vC_{x}$, $\vC_{y}$ and $\vC_{z}$ are the velocity components in the respective directions
$x$, $y$ and $z$.

\vspace{0.2cm}

The stress tensor is related to the rate of strain tensor by the constitutive or rheological
equation of the fluid which takes a differential or integral form. The rate of strain tensor
$\rsTen$ is related to the fluid velocity vector $\fVel$, which describes the steepness of velocity
variation as one moves from point to point in any direction in the flow at a given instant in time,
by the relation
\begin{equation}\label{rsTen-fVel}
    \rsTen = \nabla \fVel + (\nabla \fVel)^{\textrm{T}}
\end{equation}
where $(.)^{\textrm{T}}$ is the tensor transpose and $\nabla \fVel$ is the fluid velocity gradient
tensor defined by
\begin{equation}\label{fVelGradTen}
    \nabla \fVel =
    \left(
    \begin{array}{ccc}
      \frac{\partial \vC_{x}}{\partial x} &
      \frac{\partial \vC_{x}}{\partial y} &
      \frac{\partial \vC_{x}}{\partial z}
      \\
      \frac{\partial \vC_{y}}{\partial x} &
      \frac{\partial \vC_{y}}{\partial y} &
      \frac{\partial \vC_{y}}{\partial z}
      \\
      \frac{\partial \vC_{z}}{\partial x} &
      \frac{\partial \vC_{z}}{\partial y} &
      \frac{\partial \vC_{z}}{\partial z}
    \end{array}
    \right)
\end{equation}
with $\fVel = (\vC_{x}, \vC_{y}, \vC_{z})$. It should be remarked that the sign and index
conventions used in the definitions of these tensors are not universal.

\vspace{0.2cm}

A fluid possesses \vy\ if it is capable of storing elastic energy. A sign of this is that stresses
within the fluid persist after the deformation has ceased. The duration of time over which
appreciable stresses persist after cessation of deformation gives an estimate of what is called the
relaxation time. The relaxation and retardation times, $\rxTim$ and $\rdTim$ respectively, are
important physical properties of \vc\ fluids because they characterize whether \vy\ is likely to be
important within an experimental or observational timescale. They usually have the physical
significance that if the motion is suddenly stopped the stress decays as $e^{-t/\rxTim}$, and if
the stress is removed the rate of strain decays as $e^{-t/\rdTim}$.

\vspace{0.2cm}

For viscous flow, the \Reynolds\ number $Re$ is a dimensionless number defined as the ratio of the
inertial to viscous forces and is given by
\begin{equation}\label{Reynolds}
    Re = \frac{\rho l_{c} v_{c}}{\mu}
\end{equation}
where $\rho$ is the mass density of the fluid, $l_{c}$ is a characteristic length of the flow
system, $v_{c}$ is a characteristic speed of the flow and $\mu$ is the viscosity of the fluid.

\vspace{0.2cm}

For \vc\ fluids the key dimensionless group is the \Deborah\ number which is a measure of
elasticity. This number may be interpreted as the ratio of the magnitude of the elastic forces to
that of the viscous forces. It is defined as the ratio of a characteristic time of the fluid
$\fTim_{c}$ to a characteristic time of the flow system $t_{c}$
\begin{equation}\label{Deborah}
    De = \frac{\fTim_{c}}{t_{c}}
\end{equation}
The \Deborah\ number is zero for a \NEW\ fluid and is infinite for a \Hookean\ elastic solid. High
Deborah numbers correspond to elastic behavior and low \Deborah\ numbers to viscous behavior. As
the characteristic times are process-dependent, materials may not have a single \Deborah\ number
associated with them.

\vspace{0.2cm}

Another dimensionless number which measures the elasticity of the fluid is the \Weissenberg\ number
$We$. It is defined as the product of a characteristic time of the fluid $\fTim_{c}$ and a
characteristic strain rate $\sR_{c}$ in the flow
\begin{equation}\label{Weissenberg}
    We = \fTim_{c} \sR_{c}
\end{equation}

Other definitions to \Deborah\ and \Weissenberg\ numbers are also in common use and some even do
not differentiate between the two numbers. The characteristic time of the fluid may be taken as the
largest time constant describing the molecular motions, or a time constant in a constitutive
equation. The characteristic time for the flow can be the time interval during which a typical
fluid element experiences a significant sequence of kinematic events or it is taken to be the
duration of an experiment or experimental observation. Many variations in defining and quantifying
these characteristics do exit, and this makes \Deborah\ and \Weissenberg\ numbers not very well
defined and measured properties and hence the interpretation of the experiments in this context may
not be totally objective.

\vspace{0.2cm}

The \Boger\ fluids are constant-viscosity \nNEW\ elastic fluids. They played important role in the
recent development of the theory of fluid elasticity as they allow dissociation between elastic and
viscous effects. \Boger\ realized that the complication of variable viscosity effects can be
avoided by employing test liquids which consist of low concentrations of flexible high molecular
weight polymers in very viscous solvents, and these solutions are nowadays called \Boger\ fluids.

} 

\newpage

\section[Nomenclature]{Appendix B} \label{Nomenclature}

{\Large \bf Nomenclature}

\begin{supertabular}{ll}
  $\verb|       |$      & \\
  $\eAlpha$             & parameter in \Ellis\ model \\
  $\str$                & strain \\
  $\sR$                 & rate of strain (s$^{-1}$) \\
  $\sR_{c}$             & characteristic rate of strain (s$^{-1}$) \\
  $\rsTen$              & rate of strain tensor \\
  $\epsilon$            & porosity \\
  $\elong$              & extension \\
  $\elongR$             & rate of extension (s$^{-1}$) \\
  $\rxTimF$             & structural relaxation time in \Fredrickson\ model (s) \\
  $\rxTim$              & relaxation time (s) \\
  $\rdTim$              & retardation time (s) \\
  $\fTim_{c}$           & characteristic time of fluid (s) \\
  $\fgTim$              & first time constant in \Godfrey\ model (s) \\
  $\sgTim$              & second time constant in \Godfrey\ model (s) \\
  $\seTim$              & time constant in \SEM\ (s) \\
  $\Vis$                & viscosity (Pa.s) \\
  $\eVis$               & effective viscosity (Pa.s) \\
  $\iVis$               & initial-time viscosity (Pa.s) \\
  $\inVis$              & infinite-time viscosity (Pa.s) \\
  $\lVis$               & zero-shear viscosity (Pa.s) \\
  $\sVis$               & shear viscosity (Pa.s) \\
  $\exVis$              & extensional (elongational) viscosity (Pa.s) \\
  $\hVis$               & infinite-shear viscosity (Pa.s) \\
  $\fdVis$              & viscosity deficit associated with $\fgTim$ in \Godfrey\ model (Pa.s) \\
  $\sdVis$              & viscosity deficit associated with $\sgTim$ in \Godfrey\ model (Pa.s) \\
  $\rho$                & fluid mass density (kg.m$^{-3}$) \\
  $\sS$                 & stress (Pa) \\
  $\sTen$               & stress tensor \\
  $\hsS$                & stress when $\Vis = \lVis / 2$ in \Ellis\ model (Pa) \\
  $\ysS$                & \yields\ (Pa) \\
  $\verb|       |$      & \\
%
%
  $c$                   & dimensionless constant in \SEM\ \\
  $C$                   & consistency factor in \powlaw\ and \HB\ models (Pa.s$^{n}$) \\
  $C'$                  & tortuosity factor \\
  $De$                  & \Deborah\ number \\
  $D_{p}$               & particle diameter (m) \\
  $G$                   & geometric conductance (m$^4$) \\
  $G'$                  & flow conductance (m$^3$.Pa$^{-1}$.s$^{-1}$) \\
  $\Go$                 & elastic modulus (Pa) \\
  $\kF$                 & parameter in \Fredrickson\ model (Pa$^{-1}$) \\
  $K$                   & absolute permeability (m$^{2}$) \\
  $L$                   & length of tube or bed (m) \\
  $l_{c}$               & characteristic length of the flow system (m) \\
  $n$                   & flow behavior index \\
  $\fNSD$               & first normal stress difference (Pa) \\
  $\sNSD$               & second normal stress difference (Pa) \\
  $P$                   & pressure (Pa) \\
  $\Delta P$            & pressure drop (Pa) \\
  $q$                   & superficial (\Darcy) velocity (m.s$^{-1}$) \\
  $Q$                   & volumetric flow rate (m$^{3}$.s$^{-1}$) \\
  $R$                   & tube radius (m) \\
  $Re$                  & \Reynolds\ number \\
  $R_{eq}$              & equivalent radius (m) \\
  $t$                   & time (s) \\
  $t_{c}$               & characteristic time of flow system (s) \\
  $\Tr$                 & \TR\ ratio \\
  $\fVel$               & fluid velocity vector \\
  $\nabla \fVel$        & fluid velocity gradient tensor \\
  $v_{c}$               & characteristic speed of flow system (m.s$^{-1}$) \\
  $We$                  & \Weissenberg\ number \\
  $\verb|       |$      & \\
\end{supertabular}

\vspace{0.2cm} 

{\bf \em \noindent Abbreviations}: %
\vspace{-0.3cm} \noindent

\begin{supertabular}{ll}
  $\verb|       |$      & \\
  2D                    & two-dimensional \\
  3D                    & three-dimensional \\
  BKC                   & \Blake-\Kozeny-\Carman \\
  IPM                   & \InvPM \\
  MTP                   & Minimum Threshold Path \\
  TYP                   & Threshold Yield Pressure \\
  $(\cdot)^{T}$         & matrix transpose \\
  $\ucd \cdot$          & upper convected time derivative \\
  $\verb|       |$      & \\
\end{supertabular}

\vspace{0.5cm}

\noindent %
{\bf Note}: units, when relevant, are given in the SI system. Vectors and tensors are marked with
boldface. Some symbols may rely on the context for unambiguous identification.

\newpage
\phantomsection \addcontentsline{toc}{section}{\protect \numberline{} References} %
\bibliographystyle{unsrt}    %
\bibliography{Biblio}        %

} 

\newpage
\phantomsection \addcontentsline{toc}{section}{\protect \numberline{} Index} %
\printindex

\end{document}